\RequirePackage[2020-02-02]{latexrelease}
\documentclass[%
 reprint,
 amsmath,amssymb,
 aps,
]{revtex4-2}
\usepackage{color}
\usepackage{graphicx}% Include figure files
\usepackage{geometry}
\usepackage{dcolumn}% Align table columns on decimal point
\usepackage{bm}% bold math
\usepackage{float}
\floatstyle{plaintop}
\restylefloat{table}
\usepackage{soul}
\usepackage{caption}
\usepackage{adjustbox}
\usepackage{amsmath}

\usepackage[mathlines]{lineno}% Enable numbering of text and display math
\begin{document}

\preprint{APS/123-QED}

\title{The Rheology of Granular Mixtures with Varying Size, Density, Particle Friction and Flow Geometry}% Force line breaks with \\
%\thanks{A footnote to the article title}%
\author{Eric C.P. Breard}%
 \email{Eric.Breard@ed.ac.uk}
\affiliation{%
 School of Geosciences,\\ University of Edinburgh, Edinburgh, UK
  \\Department of Earth Sciences,\\ University of Oregon, Eugene, OR, USA
}%

\author{Luke Fullard}
\email{Luke.Fullard@horizons.govt.nz}
 \affiliation{Science and Innovation Team, \\ Horizons Regional Council, New Zealand.}
 %Lines break automatically or can be forced with \\

\author{Josef Dufek}%
 %\email{ebreard@uoregon.edu}
\affiliation{%
 Department of Earth Sciences,\\ University of Oregon, Eugene, OR, USA
}%

\date{\today}% It is always \today, today,
             %  but any date may be explicitly specified

\begin{abstract}
Employing the discrete element method, we study the rheology of dense granular media mixtures, varying in size, density, and frictional properties of particles, across a spectrum from quasi-static to inertial regimes. By accounting for the volumetric contribution of each solid phase, we find that the stress ratio, $\mu$, and concentration $\phi$, scale with the inertial number when using volume averaging to calculate mean particle density, friction and size. Moreover, the critical packing fraction correlates with skewness, polydispersity, and particle friction, irrespective of the size distribution. Notably, following the work of Kim and Kamrin [2020] we introduce a rheological power-law scaling to collapse all or monodisperse and polydisperse data, reliant on concentration, dimensionless granular temperature, and the inertial number. This model seamlessly merges the $\mu(I)$-rheology and Kinetic Theory, enabling the unification of all local and non-local rheology data onto a single master curve.

\end{abstract}

\pacs{Valid PACS appear here}% PACS, the Physics and Astronomy
                             % Classification Scheme.
%\keywords{Suggested keywords}%Use showkeys class option if keyword
                              %display desired
\maketitle

%\tableofcontents

\section{\label{sec:Intro}Background and Introduction}
Since granular flows are ubiquitous, understanding the flowing behaviour of grains and powders has been increasingly necessary to predict hazards related to geophysical flows (e.g. snow avalanches \cite{sovilla2018intermittency}, debris flows \cite{iverson2011}, rock avalanches \cite{hungr2006rock}, pyroclastic density currents \cite{lube2020multiphase,dufek2016fluid,cole2015hazards}), to understand processes on other planetary bodies (e.g. Mars \cite{huber2020physical}, Titan \cite{mendez2017electrification}), and to support applications in multi-billion pound/dollar industries and governmental agencies throughout the world \cite{andreotti2013granular} (e.g. pharmaceutical, agricultural, food industries, US Departments of Energy and Defense). 
Depending on particle shape, stiffness, friction and size distribution, granular flows can display for instance history dependent effects, hysteresis, non-locality, dilatancy, and coupling with the interstitial fluid, which makes them highly complex granular media \cite{andreotti2013granular,herrmann2013physics}. 
Numerous attempts have been made to build constitutive models describing granular media, such as the $\mu\left ( I \right )$-rheology that suggests a one-to-one relationship between the inertial number defined as $I\equiv \dot{\gamma }{\bar{d}}/\sqrt{P/\rho_s}$ and the shear-to-normal stress ratio $\tau/P$, where $\tau$ is the shear stress, $P$ is the normal stress, $\bar{d}$ is the mean particle diameter, $\dot{\gamma }$ is the shear rate and $\rho_s$ is the mean particle density \cite{jop2006constitutive,gdr2004dense,staron2014continuum}.
Most studies on granular flows focus on (quasi) monodisperse (single particle size) and monophasic (e.g. \cite{fullard2018testing}) mixtures of grains where all grains have the same frictional properties and densities, whereas natural mixtures can be made of particles with various densities, and friction properties and are polydisperse (e.g.\cite{breard2019}). It is unclear at this time whether these complexities can be captured by a single rheological model \cite{iverson2003debris}.
In this work we first investigate the rheology of granular flows in a simple shear setup using the Discrete Element Method (DEM) and performed simulations with monodisperse, polydisperse and biphasic distributions (varying particle densities and particle-particle friction coefficients), thus expanding the work from Gu et al. \cite{gu2017}. First, we demonstrate the quasi-static, intermediate and inertial-collisional regimes persist for all mixtures, wherein solid concentration and stress ratio scale with the inertial number when taking into account the volumetric mean $D_{43}$. Second, following the recent work by Kim and Kamrin \cite{KK2020} we attempt to link the Kinetic Theory to the $\mu\left ( I \right )$-rheology through the dimensionless granular temperature $\Theta\equiv \rho_sT/P$, where T is the granular temperature. Finally, we unify the rheology of the granular mixtures on various flow geometry using a modified power-law scaling $\mu(I,\Theta,\phi)$, demonstrating the need to account for the concentration to collapse our data on a master curve.

\section{\label{sec:Methods}Materials and Methods}
\subsubsection{\label{secsub:DEM}Discrete Element Method}
\begin{figure*}
\includegraphics{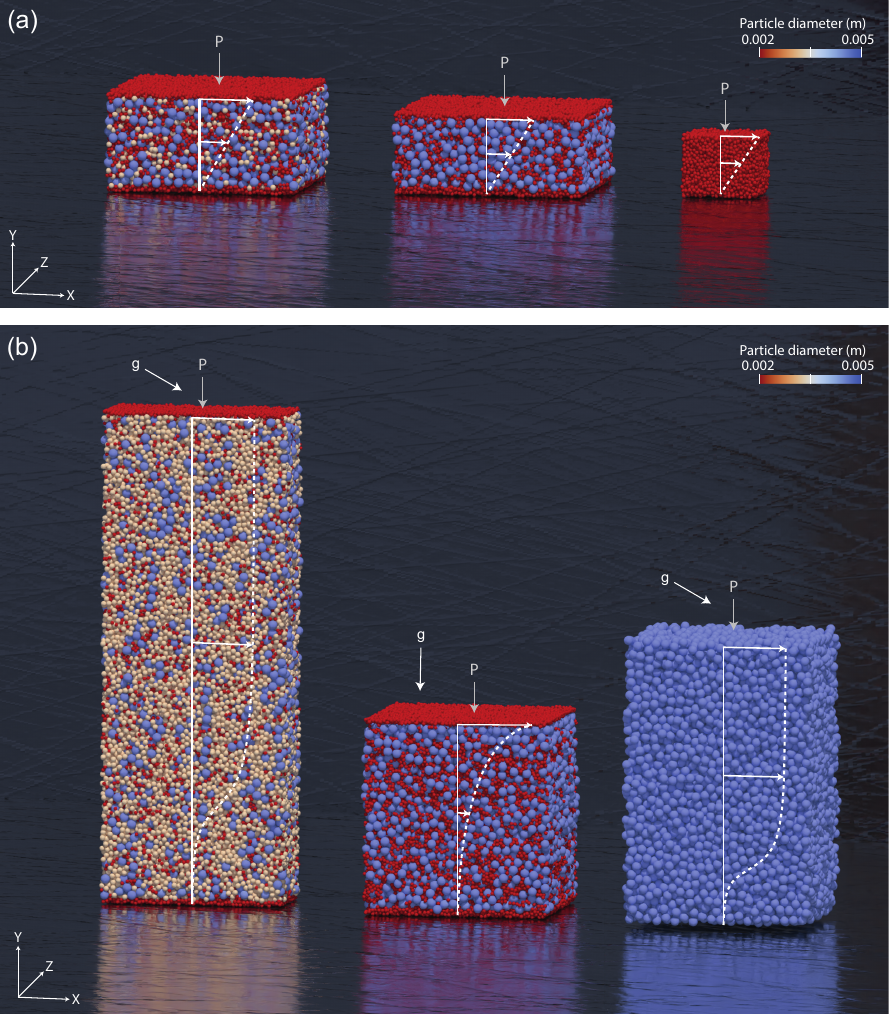}
\caption{Planar shear geometry for local and non-local rheology tests. (a) Simple shear for tridisperse, bidisperse and monodisperse particle size distributions (left to right) devoid of gravity field. (b) Concave flows created using a $60^\circ$ slope with gravity. All planar shear tests consist of an imposed confining pressure of 2 KPa and imposed velocity Vx of 0.0001 m/s to 32 m/s.}
\label{fig:1}
\end{figure*}

To simulate the granular flows in simple shear we used the open-source code MFIX developed by the U.S. Department of Energy \cite{garg2012} The position and momentum of particles are explicitly described according to Newton's laws:
\begin{eqnarray}
\frac{d X^{(i)}(t)}{d t}=V^{(i)}(t)\\
m^{(i)} \frac{d V^{(i)}(t)}{d t}=F_{T}^{(i)}(t)=m^{(i)} g+\\
F_{d}^{(i \in k, m)}(t)+F_{c}^{(i)}(t) \\
I^{(i)} \frac{d \Omega(t)}{d t}=T^{(i)}(t)
\end{eqnarray}
where $X^{(i)}$ is the particle position of the i-th particle within
the domain at time t, $V^{(i)}$ is the velocity, $\Omega(t)$ is the angular velocity of the i-th particle, $m^{(i)}$ is the particle mass, $g$ is gravity. $F_{c}^{(i)}$ is
the net contact force, $F_{T}$ is the sum of the forces acting on
particle i-th, $F_{d}^{(i \in k, m)}$ is the total (viscous and pressure) drag force acting on particle i, if the m-th solid phase is located within the k-th cell. $T^{(i)}$ is the sum of all torques acting on the i-th particle and $I^{(i)}$ is the moment of inertia.
Particle contacts are modelled using the soft sphere method, which uses the spring-dashpot approach that has been rigorously validated in a series of studies \cite{garg2012,Li2012}. In this soft-sphere approach, the overlap between particles is simulated by a series of springs and dashpots in the normal and tangential directions. The dashpot is used to model the loss of kinetic energy during inelastic collisions, while the spring models the rebound of a particle which is in contact with another. Both dashpot and spring are described with dampening and stiffness coefficients in both the tangential and normal directions. 
In the DEM, each solid phase is described by a distinct diameter and density. Unless specified, particles have a density of 1050 $kg/m^3$ and particle-particle friction of 0.53, motivated by values from geophysical flows. The restitution coefficient has been shown to have a negligible impact on stresses Gu et al. \cite{gu2017} and was set at 0.6 between all phases.  
Particle assemblies are described as follows:
(1) monodisperse
(1a) with variable solid density (500 $kg/m^3$, 1050 $kg/m^3$ and 2500 $kg/m^3$)
(1b) with a diameter of 1.75 mm, 2 mm, 3 mm, 4 mm and 5 mm
(1c) with particle friction of (0.09, 0.25, 0.8)
(1d) density biphasic (density pairs of 1050-2500 $kg/m^3$, 500-1050 $kg/m^3$ and 500-2500 $kg/m^3$
(1e) friction biphasic (0.09-0.53, 0.25-0.53, 0.53-0.8 and 0.25-0.8)
(2) polydisperse
(2a) bidisperse (1.75-5 mm, 2-5 mm, 3-5 mm and 4-5 mm)
(2b) tridisperse (5-3-2 mm and 5-4-2 mm).
Each size distribution can be analyzed using the polydispersity parameter based on their radii r:
$\delta =\frac{\sqrt{\left \langle \Delta r^{2} \right \rangle}}{\left \langle r \right \rangle}$
and the skewness parameter:
$S=\frac{\left \langle \Delta r^{3} \right \rangle}{\left \langle \Delta r^{2}  \right \rangle^\frac{3}{2}}$, which measures the spread and shape of the PSD, respectively \cite{gu2017}. In the equations, $\Delta r=r-\left \langle \Delta r \right \rangle$ and the moment of $\Delta r$ and r and are defined as $\left \langle \Delta r^{n} \right \rangle=\int \Delta r^{n}P(r)dr$ and $\left \langle r^{n} \right \rangle=\int r^{n}P(r)dr$, respectively
The mean particle diameter chosen as the volume-mean diameter $(D_{43})$ is defined as follows:
\begin{eqnarray}
D_{43} \equiv \frac{\sum_{i=1}^{N}d_i^{4}}{\sum_{i=1}^{N}d_i^{3}}
\end{eqnarray}
where N is the total number of particles and $d_i$ is the diameter of particle $i$.
In our simulations, the particle stiffness is set at $k_{n}=10^{4}Pd$ to ensure particle interactions fall in the hard contact regime \cite{zhang2017microscopic}. 
Assemblies of grains were placed in a periodic box with fixed volume and vertically bounded by a rough bottom static plate and a top rough plate that imposed a confining pressure and moved at a set velocity in one direction. Roughness is generated by fixing the relative position of particles based in a monolayer. A second rough plate of particles was created away from the plates to prevent slip at high top plate velocities and to ensure a constant shear rate across the bed in the simple shear simulations (Figure \ref{fig:1}a). Each simulation consists of an initial phase where the bed is compacted and pre-sheared, followed by a phase where the top plate reaches the set velocity and that velocity is held constant. Once it reaches a steady state, the particle velocity, contact forces, diameter, density, and friction are exported and used for post-processing using the coarse-graining (CG) method.
For each set, 22 simulations were run at constant confining pressure of 2 kPa and varying top plate velocities (from 0.0001 m/s to 32 m/s) to span the quasistatic to inertial-collisional flow regimes. The simple shear simulations were conducted without the presence of a gravity field. In addition, a series of simulations were performed using different particle size distributions: monodisperse (5mm), bidisperse (2-5mm), and tridisperse (2-3-5mm). These simulations were carried out on various geometries: (a) a sloped surface with the influence of gravity, resulting in concave flows, and (b) simple shear with gravity, resulting in an exponential-like velocity profile across the bed (Figure \ref{fig:1}b), thereby inducing non-local behavior. Similar to the gravity-free simple shear simulations, only data from the flow at steady state was analyzed.

\subsubsection{\label{secsub:CoarseGraining}Coarse Graining}

The formulation of accurate continuum models of granular flows requires the need to use experimental or numerical data obtained from flows, which are discrete \cite{weinhart2016influence,goldhirsch2010stress,weinhart2013coarse}. To this end, micro–macro transition methods are used to obtain continuum fields (such as density, momentum, stress) from discrete data of individual elements (positions, velocities, orientations, interaction forces). The coarse-graining method allows us to calculate continuum fields by applying a local smoothing kernel, coarse-graining function, with a well-defined smoothing length, i.e. coarse-graining scale, that automatically generates fields satisfying the continuum equations \cite{tunuguntla2017comparing}. In this study, we use the Lucy coarse-graining function:

\begin{equation}
\begin{split}
\psi(\mathbf{r}) =\frac{105}{16 \pi c^{3}}\left[-3\left(\frac{a}{c}\right)^{4} \right. +8\left(\frac{a}{c}\right)^{3} \\
\left. -6\left(\frac{a}{c}\right)^{2}+1  \right],\text { if } 
a=\frac{|\mathbf{r}|}{c}<1, \text { else } 0,
\end{split}
\end{equation}

where the cutoff lengthscale $c=2*D_{43}$ and the coarse-graining scale $w=0.75*D_{43}$. We ran the coarse-graining analysis away from the boundaries, with an offset of $1.5*c*w + 0.5*d_{max}$. From the coarse-graining analysis, we calculate the macroscopic stress tensor:

\begin{eqnarray}
\sigma(r, t)=\sigma^{k}(r, t)+\sigma^{c}(r, t)
\end{eqnarray}

where the kinetic tensor depends on particle velocity fluctuations and the contact tensor depends on the contact forces. The full description of the tensors is provided in Breard et al. \cite{breard2022investigating}.

The granular friction coefficient is calculated from the $2 \mathrm{D}$ version of the stress tensor in combination with the pressure using all 3 contributions, which is the more appropriate approach for plane shear flow setups \cite{weinhart2013coarse}.

\begin{eqnarray}
\mu=\frac{\left|\sigma^{D}\right|}P
\end{eqnarray}

where
\begin{eqnarray}
\left|\sigma^{D}\right|=\sqrt{0.5 \sigma_{i j}^{\prime D} \sigma_{i j}^{\prime D}}
\end{eqnarray}

All the data presented in this manuscript are for beds that are yielding (i.e. where a shear rate is measurable).
The average normal stress or solid pressure in the system is calculated as the trace of the $3 \mathrm{D}$ stress tensor:
\begin{eqnarray}
P=\frac{1}{3} \operatorname{tr}(\sigma) \label{eqn:Ptrace}
\end{eqnarray}

The granular temperature inside the mixture is defined as:
\begin{eqnarray}
T_{g}=\frac{\operatorname{tr}\left(\sigma^{k}\right)}{3 \rho} \label{eqn:Tgtrace}
\end{eqnarray}

For all planar shear simulations, with and without gravity, the data presented in the manuscript is obtained by spatial and temporal averaging of the CG fields across the whole bed, excluding the cells located at a distance of three mean particle diameter from the boundaries to avoid any of their effects. In the case where we simulated the flow on a slope (concave velocity profile), we analyse with CG the bottom half of the flow where strong shear gradient exist.

\subsubsection{\label{secsub:Fitting}Empirical function fitting}
Once the flows reached steady state, we exported discrete data at 10 Hz that were post-processed using the CG method to calculate instantaneous fields. For each simulation,  the fields were space and time-averaged over a time-window of 2 to 4 seconds. 

We show the standard $\mu(I)$ and a common form of $\phi(I)$ hold in all simple shear numerical experiments, but with the fitting parameters dependent on the particle size, density, and surface friction dependence. 

\begin{eqnarray}
\mu &=& \mu_1 + \dfrac{\mu_2 - \mu_1}{1 + I_0/I}, \label{eqn:muI}\\
\phi &=& \phi_c - aI^\gamma. \label{eqn:phiI}
\end{eqnarray}

\section{\label{sec:Results}Results}
\subsection{\label{secsub:MonoResults}Monodisperse experiments in simple shear}

\subsubsection{Monodisperse simple shear, varying particle size.}
First we perform a series of simple shear cell experiments with constant particle size to find the particle size dependence of the parameters $\mu_1, \mu_2, I_0$ in Equation \ref{eqn:muI}, and $\phi_c, a, \gamma$ in Equation \ref{eqn:phiI}. 

\begin{figure}[htbp]
	\centering
		\includegraphics[width=0.45\textwidth, angle=0]{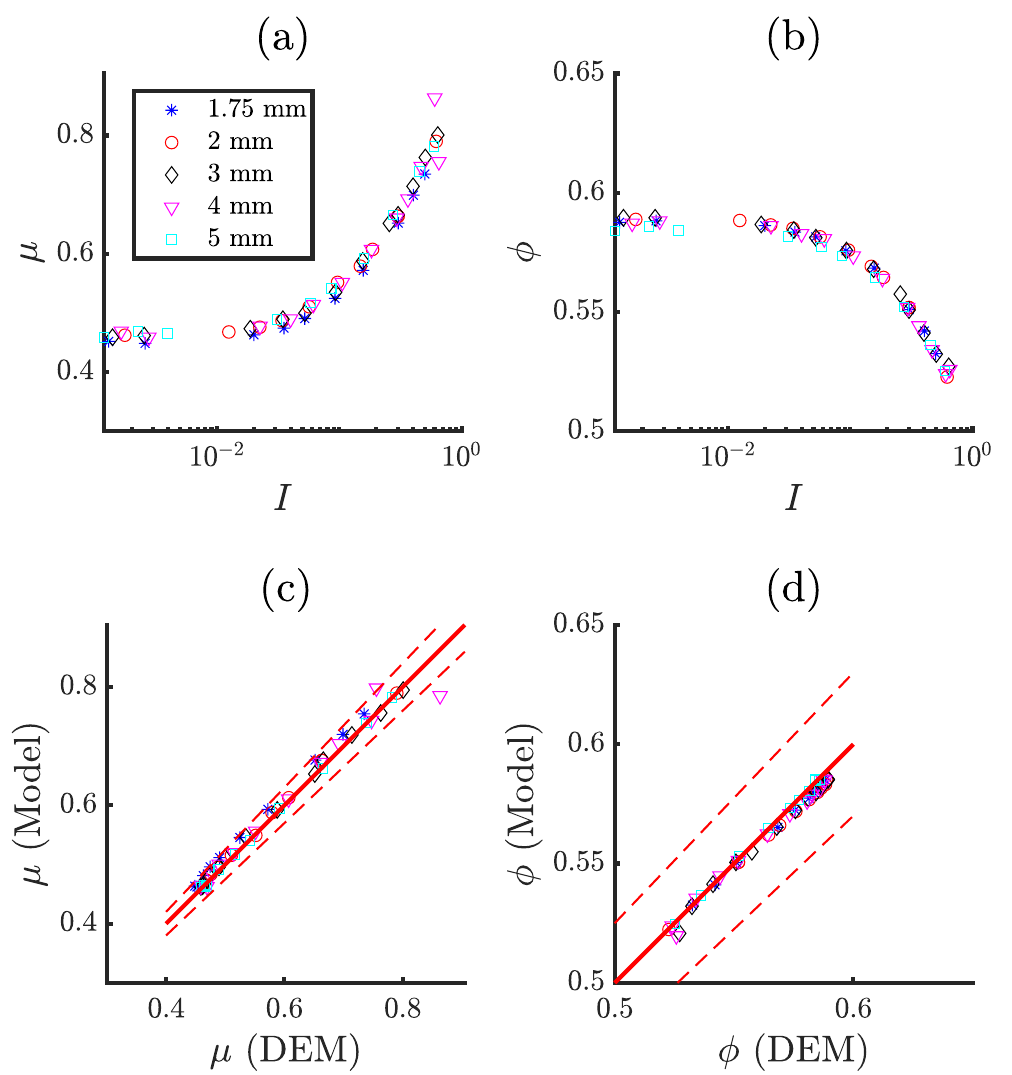}
	\caption{Stress ratio $\mu$ (a) and concentration $\phi$ (b) as a function of the inertial number $I$ for monodisperse mixtures. Modeled stress ratio (c) and concentration (d) against numerical experiments. Dotted red lines are the 95\% confidence interval for the linear fitting.}
	\label{fig:2}
\end{figure}

The parameters were found to be independent of particle size  (Note this is only considered valid in the ranges of particle sizes tested) all rounded to 3dp. The fit was conducted with the $5 mm$ data and the fitted model parameters were used to validate the size-independent assumption in Figure \ref{fig:2} .
For the friction equation, the fitted parameters are:
\begin{eqnarray}
\mu_1 &=& 0.4607, \label{eqn:MDmu1}\\
\mu_2 &=& 1.1223, \label{eqn:MDmu2}\\
I_0 &=& 0.6253, \label{eqn:MDI0}
\end{eqnarray}

For the volume fraction, the fits are:
\begin{eqnarray}
\phi_c &=& 0.5858, \label{eqn:MDphic}\\
a &=& 0.0939, \label{eqn:MDa}\\
\gamma &=& 0.8180, \label{eqn:MDgamma}
\end{eqnarray}

The collapse of stress ratio and concentration data with the Inertial number is in line with the classic $\mu\left ( I \right )$-rheology \cite{jop2006constitutive,fall2015dry}. 

\subsubsection{Simple shear, varying particle density.}
\begin{figure}[htbp]
	\centering
		\includegraphics[width=0.45\textwidth, angle=0]{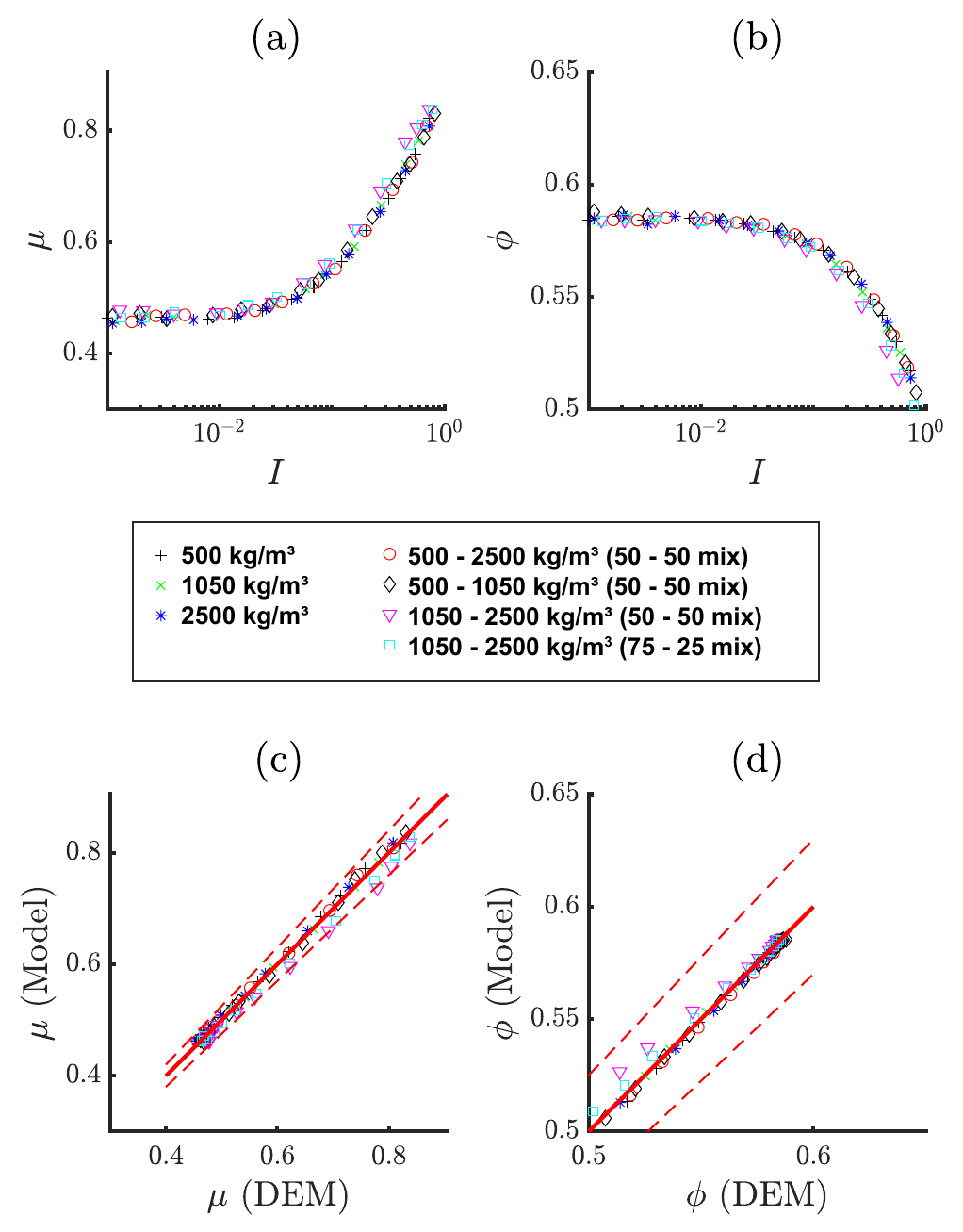}
	\caption{Stress ratio $\mu$ (a) and concentration $\phi$ (b) as a function of the inertial number $I$ for density mixtures. Modeled stress ratio (c) and concentration (d) against numerical experiments. The dotted red lines are the 95\% confidence interval for the linear fitting.}
	\label{fig:3}
\end{figure}

To investigate the rheology of bi-phasic mixtures where particles have a monodisperse grain-size distribution but varying particle density, we perform simple shear cell numerical experiments. For experiments with a constant particle density, or density mixtures, the $\mu(I)$ and $\phi(I)$ curves all fall onto the same curve (Figure \ref{fig:3}), as long as the volume-weighted density is used in the inertial number calculation. 

If $\epsilon$ is the volume weighting of the particle phase with density $\rho = \rho_1$ in the particle mixture, then $\rho_{mix} = \epsilon\rho_1 + (1-\epsilon)\rho_2$.
\subsubsection{Simple shear, varying particle surface friction coefficient.}
Simple shear numerical experiments were performed with (a) particles of the same size and density, but varying particle-particle friction coefficients and (b) particles of the same density but different diameters. The particle-particle friction coefficient varied from values as low as 0.09 up to 0.8.

\begin{figure}[htbp]
	\centering
		\includegraphics[width=0.45\textwidth, angle=0]{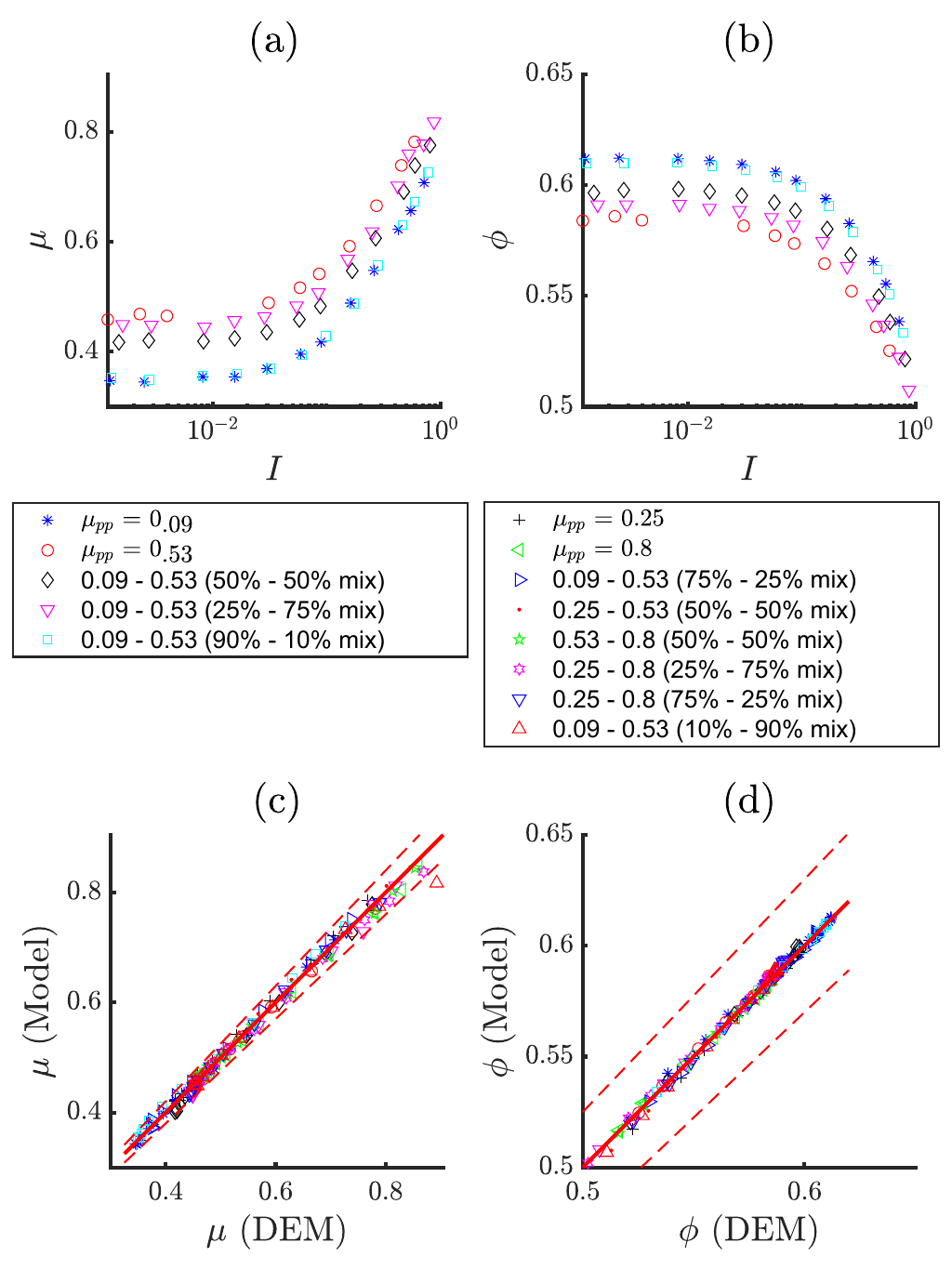}
	\caption{Stress ratio $\mu$ (a) and concentration $\phi$ (b) as a function of the inertial number $I$ for mixtures with varying particle-particle friction $\mu_{pp}$. Modeled stress ratio (c) and concentration (d) against numerical experiments. Dotted red lines are the 95\% confidence interval for the linear fitting.}
	\label{fig:4}
\end{figure}

As expected, the stress ratio $\mu$ and $\phi$ are strongly dependent on the particle-particle friction coefficient $\mu _{pp}$ (Figure \ref{fig:4}) and can be scaled with the Inertial number in the following way:
\begin{eqnarray}
\mu_1^{pp} = 0.4674 - 0.2199*\\exp(-6.1427*\mu_{pp}), \label{eqn:mono_fricmu1} \\
\mu_2^{pp} = 1.0806, \label{eqn:mono_fricmu2} \\
I_0^{pp} = 0.7306 -0.2701*\mu_{pp}, \label{eqn:mono_fricI0} 
\end{eqnarray}
and 
%       mu1:  y = 0.46738 - 0.2110.*exp^(-6.1427.*mu_p) 
%       mu2:    y = 1.0806 
%       I0:  y = -0.2702*mu_p + 0.7306

%       phi_c:   y = 0.58259 + 0.0457*e^(-4.4828*x)
%       a:  y = 0.0952
%       gamma:   y = -0.2098*x + 0.9240
\begin{eqnarray}
\phi_c^{pp} = 0.5826 + 0.0457*\\exp(-4.4828*\mu_{pp}), \label{eqn:mono_fricphic} \\
a^{pp} = 0.0952 \\
\gamma^{pp} = 0.9249 -0.2098*\mu_{pp} 
\end{eqnarray}

\begin{figure*}[htbp]
	\centering
		\includegraphics[width=0.8\textwidth, angle=0]{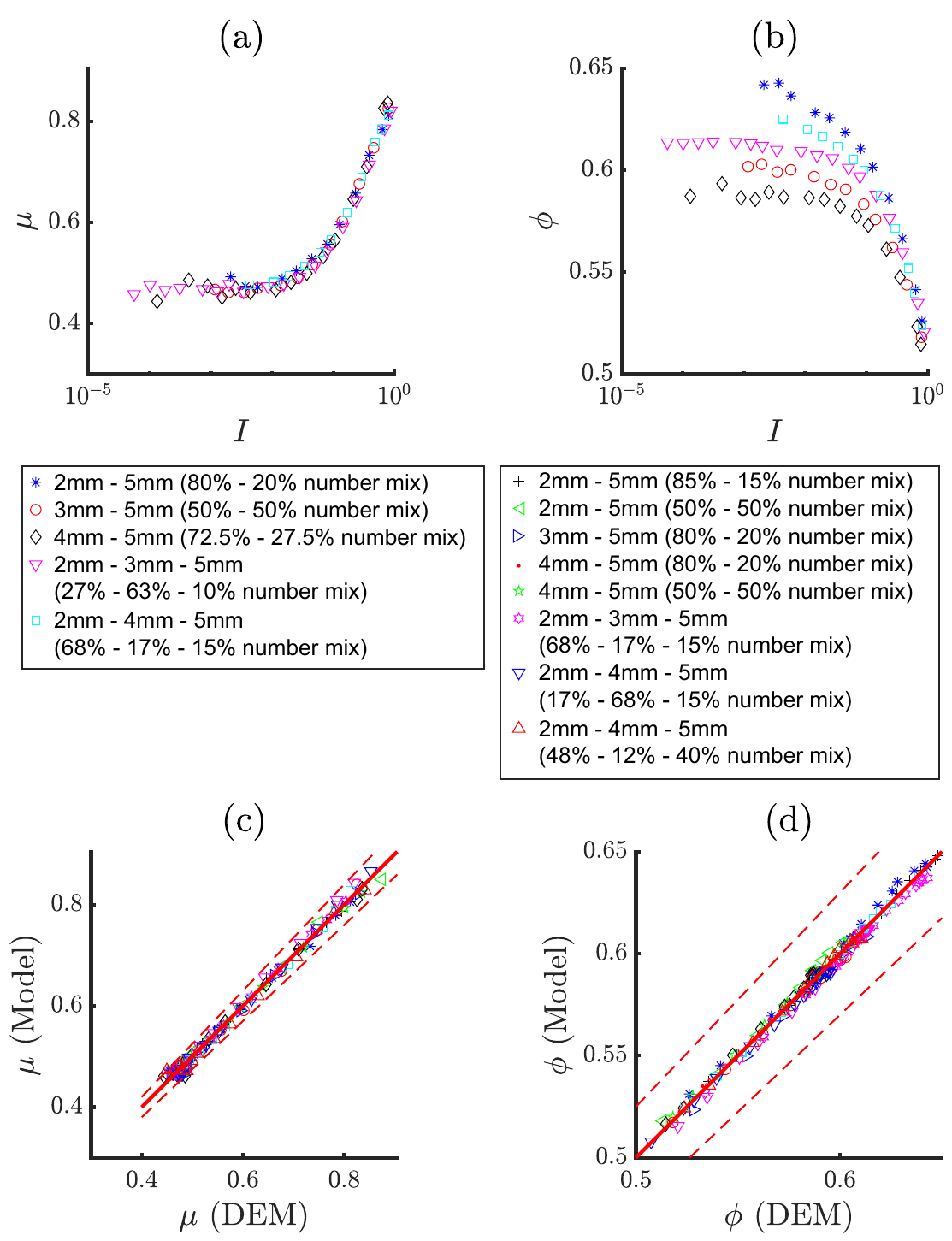}
	\caption{Stress ratio $\mu$ (a) and concentration $\phi$ (b) as a function of the inertial number $I$ for polydisperse size mixtures. Modeled stress ratio (c) and concentration (d) against numerical experiments. The dotted red lines are the 95\% confidence interval for the linear fitting.}
	\label{fig:5}
\end{figure*}

The $\mu(I)$ coefficients for the particle mixtures were found to be volumetrically related to the individual low ($L$) and high ($H$) coefficients. That is
\begin{align}
\mu_1^{mix} & = (1-\epsilon)\mu_1^H + \epsilon\mu_1^L, \label{eqn:fricmu1} \\
\mu_2^{mix} & = (1-\epsilon)\mu_2^H + \epsilon\mu_2^L, \label{eqn:fricmu2} \\
I_0^{mix} & = (1-\epsilon)I_0^H + \epsilon I_0^L, \label{eqn:fricI0} 
\end{align}
For the particle mixtures the $\phi(I)$ coefficient $\phi_c$ was found to be volumetrically related to the individual low ($L$) and high ($H$) coefficients, but the $a$ and $\gamma$ coefficients were not found to be a function of particle friction (Figure \ref{fig:5}). Instead, they are only a function of particle diameter, as expressed in Equations \ref{eqn:MDa} and \ref{eqn:MDgamma}. That is:

\begin{eqnarray}
\phi_c^{mix} = (1-\epsilon)\phi_1^H + \epsilon\phi_1^L, \label{eqn:fricphic} \\
a^{mix} = a^H = a^L, \label{eqn:a} \\
\gamma^{mix} = \gamma^H = \gamma^L, \label{eqn:muI0} 
\end{eqnarray}

\subsection{\label{secsub:MixResults}Polydisperse experiments in simple shear}
As demonstrated by \cite{gu2017}, the friction and dilatancy of polydisperse granular-size mixtures are a function of the
Skewness and Polydispersity parameters (see Methods section for definition). In addition, we define $\epsilon_{S/M/L}$ as the volume fraction of the small/medium/large phase.
Therefore, the friction law coefficients of Equation \ref{eqn:muI} can be written as follows:

\begin{eqnarray}
\mu_1^{mix} &=& \left(\sum_i \epsilon_i\mu_1^i\right) + C^1_1 + C^1_2\delta + C^1_3S\delta^2, \label{eqn:polymu1} \\
\mu_2^{mix} &=& \left(\sum_i \epsilon_i\mu_2^i\right) + C^2_1 + C^2_2\delta + C^2_3S\delta^2, \label{eqn:polymu2} \\
I_0^{mix} &=& \left(\sum_i \epsilon_iI_0^i\right) + C^3_1 + C^3_2\delta + C^3_3S\delta^2, \label{eqn:polymuI0} 
\end{eqnarray}
where $C^i_{1,2,3}$ are constants (Table \ref{Table:1}).

\begin{table}[]
\begin{tabular}{ll|l|l|l|}

\cline{3-5}
                              &         & $C^i_1$                & $C^i_2$    & $C^i_3$  \\ \hline
\multicolumn{1}{|l|}{$i = 1$} & $\mu_1$ & $-0.002243$            & $0.03100$  & $0.003543$  \\ \hline
\multicolumn{1}{|l|}{$i = 2$} & $\mu_2$ & $-0.09007$             & $0.3527$   & $-0.7073$ \\ \hline
\multicolumn{1}{|l|}{$i = 3$} & $I_0$   & $-0.1926$               & $0.6418$    & $-1.040$ \\ \hline
\multicolumn{1}{c}{} &\multicolumn{1}{c}{} & \multicolumn{1}{c}{} & \multicolumn{1}{c}{} & \multicolumn{1}{c}{} \\
\multicolumn{1}{c}{} &\multicolumn{1}{c}{} & \multicolumn{1}{c}{} & \multicolumn{1}{c}{} & \multicolumn{1}{c}{} \\
\cline{3-5}
                              &          & $D^i_1$    & $D^i_2$ & $D^i_3$  \\ \hline
\multicolumn{1}{|l|}{$i = 1$} & $\phi_c$ & $0.0002977$ & $0.05430$ & $0.1195$  \\ \hline
\multicolumn{1}{|l|}{$i = 2$} & $a$      & $-0.007443$  & $0.04073$   & $0.08477$  \\ \hline
\multicolumn{1}{|l|}{$i = 3$} & $\gamma$ & $0.006517$   & $-0.5353$   & $-0.1435$ \\ \hline
\caption{Fitting parameters for the polydisperse mixtures.}\label{Table:1}
\end{tabular}
\end{table}

%phicmixer = 0.5857    +   0.000297706585696006  + 0.0543022481670068.*Polydispersity(iiii) + 0.119458313573459.*Skewness(iiii).*Polydispersity(iiii).^2; 
         %amixer = 0.0939       +    -0.00744342495899381  + 0.0407336100276659 .*Polydispersity(iiii) + 0.0847671355209712.*Skewness(iiii).*Polydispersity(iiii).^2;
         %gammamixer = 0.8180    +    0.00651657405842254  + -0.535296304497962  .*Polydispersity(iiii) + -0.14354112741381.*Skewness(iiii).*Polydispersity(iiii).^2;
         
Similarly, the dilation law (Equation \ref{eqn:phiI}) can be written to account for the particle size distribution of the granular mixture:
\begin{eqnarray}
\phi_c^{mix} &=& \left(\sum_i \epsilon_i\phi_c^i\right) + D^1_1 + D^1_2\delta + D^1_3S\delta^2, \label{eqn:polyphi1} \\
a^{mix} &=& \left(\sum_i \epsilon_ia^i\right) + D^2_1 + D^2_2\delta + D^2_3S\delta^2, \label{eqn:polyphi2} \\
\gamma^{mix} &=& \left(\sum_i \epsilon_i\gamma^i\right) + D^3_1 + D^3_2\delta + D^3_3S\delta^2, \label{eqn:polyphiI0} 
\end{eqnarray}
where $D^i_{1,2,3}$ are constants presented in Table \ref{Table:1}.
%\begin{table}[]
%\begin{tabular}{ll|l|l|l|}
%\cline{3-5}
%                              &          & $D^i_1$    & %$D^i_2$ & $D^i_3$  \\ \hline
%\multicolumn{1}{|l|}{$i = 1$} & $\phi_c$ & $0.0002977$ %& $0.05430$ & $0.1195$  \\ \hline
%\multicolumn{1}{|l|}{$i = 2$} & $a$      & $-0.007443$  %& $0.04073$   & $0.08477$  \\ \hline
%\multicolumn{1}{|l|}{$i = 3$} & $\gamma$ & $0.006517$   %& $-0.5353$   & $-0.1435$ \\ \hline
%\end{tabular}
%\end{table}

\section{\label{sec:Discussion}Discussion}
In the present study, we explored the biphasic mixtures of grains with different particle-particle friction coefficients, densities and size distributions and found the Inertial number is indeed capable of capturing the evolution of the stress ratio $\mu$ and dilatancy $\phi$ provided that one account for the contribution of each particle phase by volume using the D[4,3] as the characteristic particle lengthscale in the system. Although our research is limited to particles within the hard contact regime, it carries significant relevance for the majority of both natural and industrial granular flows.
In the past decades, the Kinetic Theory (KT) originally developed for gases, has been adapted to describe the frictional regime of granular flows \cite{berzi2020extended,lun1984kinetic,garzo1999dense,jenkins1983theory}, but this modified KT has yet to be adapted to describe polydisperse mixtures. Expanding on the Kinetic Theory (KT) that characterizes the solid pressure in relation to the granular temperature, Kim and Kamrin \cite{KK2020, kim2023} have recently implemented a power-law scaling for the stress ratio. This scaling correlates with the Inertial number and the dimensionless granular temperature, defined as $\Theta =\rho {s}T{g}/P$. Granular temperature has proven to be a critical parameter in the development of constitutive models for flows that exhibit both liquid- and solid-like behaviors \cite{zhu2023granular}.

\begin{figure*}[htbp]
	\centering
		\includegraphics[width=0.95\textwidth, angle=0]{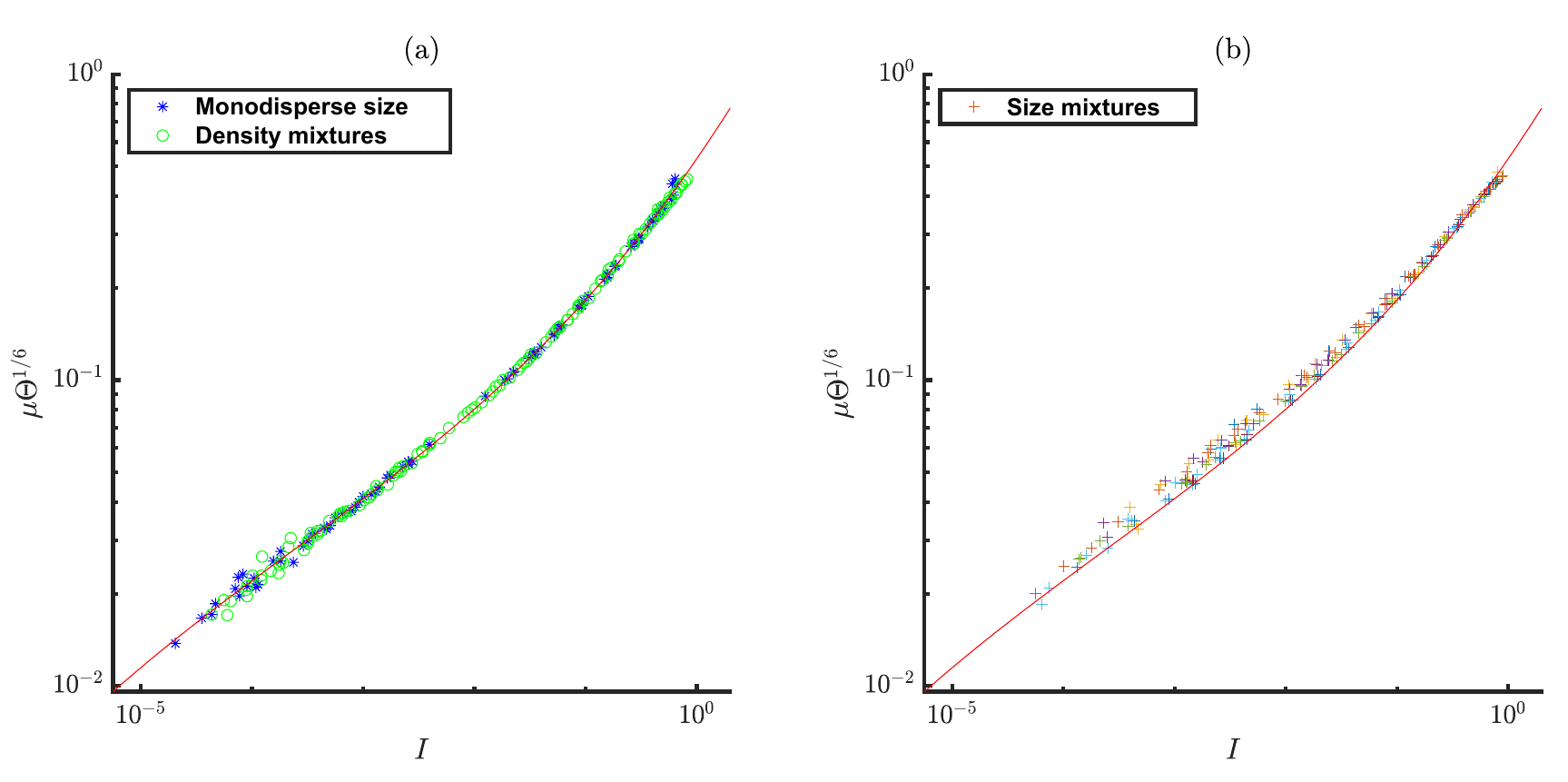}
	\caption{Power-law scaling as defined by Kim and Kamrin [cite] $\mu \Theta ^{1/6}=f(I)$ for monodisperse and density mixtures (a) and size mixtures (b)  }
	\label{fig:6}
\end{figure*}

In light of these findings, we investigate a potential scaling $\mu(I,\Theta)$ across all the size distributions. We observe that "heating" the material (endogenous mechanical vibrations), or increasing the granular temperature, tends to soften the material and it follows that $\mu\Theta^{p}=f(I)$ where p equals 1/6 (as shown in Figure \ref{fig:6}a).

Contrary to Kim and Kamrin's approach \cite{KK2020}, which relied solely on the streamwise velocity component to calculate granular temperature and the vertical component to calculate solid pressure, we employed all three components of stress (refer to Equations \ref{eqn:Ptrace} and \ref{eqn:Tgtrace}).

Our data is aptly represented by the following polynomial expression:
\begin{align}
\ln\left(\mu\Theta^{1/6} \right) = & 0.0013 \ln^3\left(I\right) + 0.0315 \ln^2\left(I\right) \nonumber \\
& + 0.5276\ln\left(I\right) - 0.6357 \label{eqn:kamrin_fluidity_fit_original}
\end{align}

While the monodisperse and density mixture data falls perfectly on the same curve, the data from polydisperse simulations does not coincide with the monodisperse data across the spectrum of quasi-static to inertial/collisional regimes (see Figure \ref{fig:6}b).

\begin{figure*}[htbp]
	\centering
		\includegraphics[width=0.95\textwidth, angle=0]{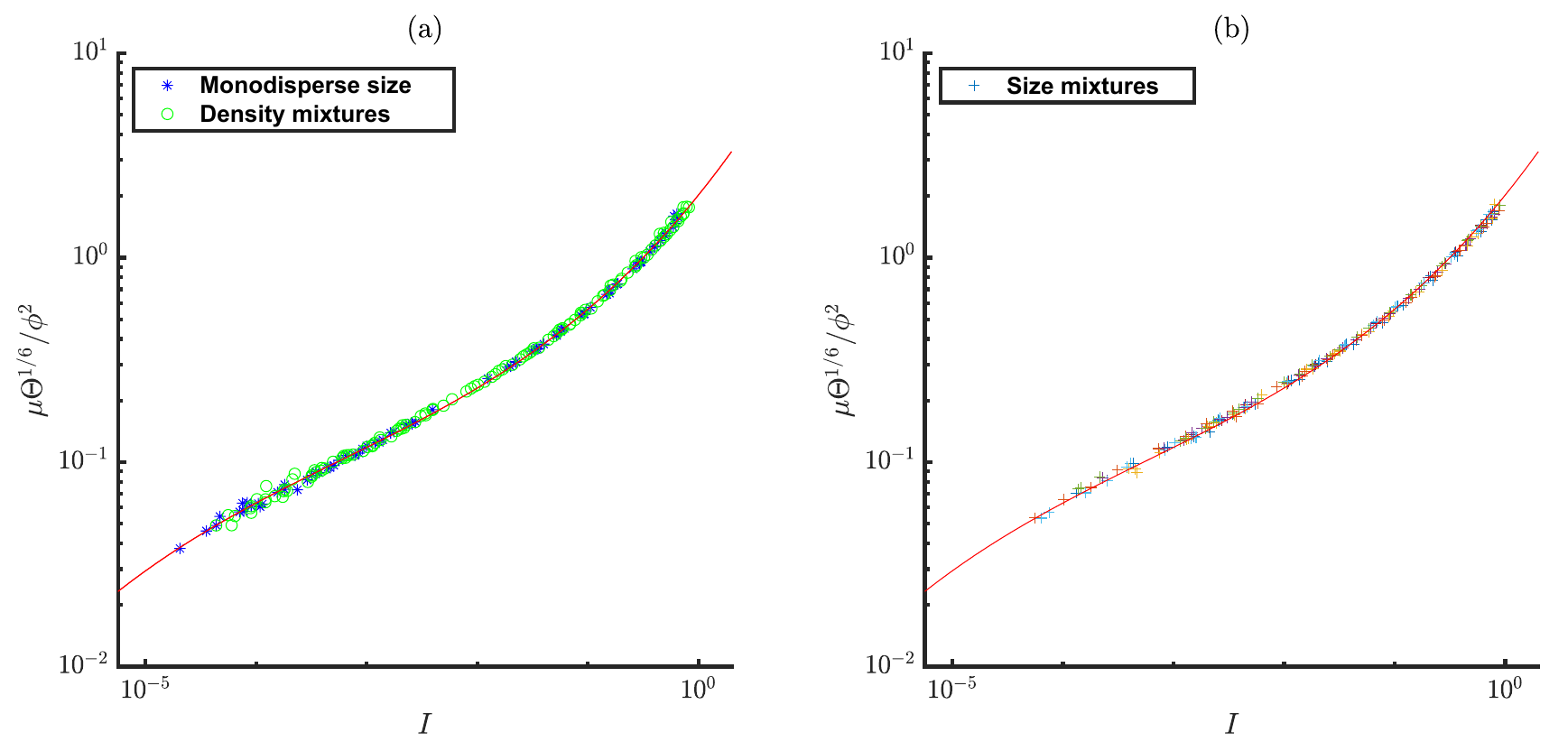}
	\caption{Modified Power-law scaling $\frac{\mu \Theta ^{1/6}}{\phi^{2}} = f(I)$ for monodisperse and density mixtures (a) and size mixtures (b). The red line is the same on both plots}
	\label{fig:7}
\end{figure*}

In our pursuit to unify the data, we explore alternative approaches. The conventional scaling $\mu(I,\Theta)$, as demonstrated by Kim (2023) \cite{kim2023}, effectively eliminates the rheological dependence on $\phi$, but it does not apply to polydisperse mixtures since $\phi$ and $\mu$ data does not collapse due to the polydispersity (\ref{Supplementary fig:1}). Nevertheless, we first can establish a modified power-law scaling, $\mu(I,\Theta,\phi)$, fitted exclusively for the monodisperse data. The ensuing formulation is presented in Figure \ref{fig:7}a:

\begin{align}
\ln\left(\mu\Theta^{1/6}/\phi^2 \right) = & 0.0024 \ln^3\left(I\right) \nonumber + 0.0543 \ln^2\left(I\right) \nonumber \\
& + 0.6704\ln\left(I\right) + 0.7036 \label{eqn:kamrin_fluidity_fit}
\end{align}

\begin{figure}[htbp]
	\centering
		\includegraphics[width=0.4\textwidth, angle=0]{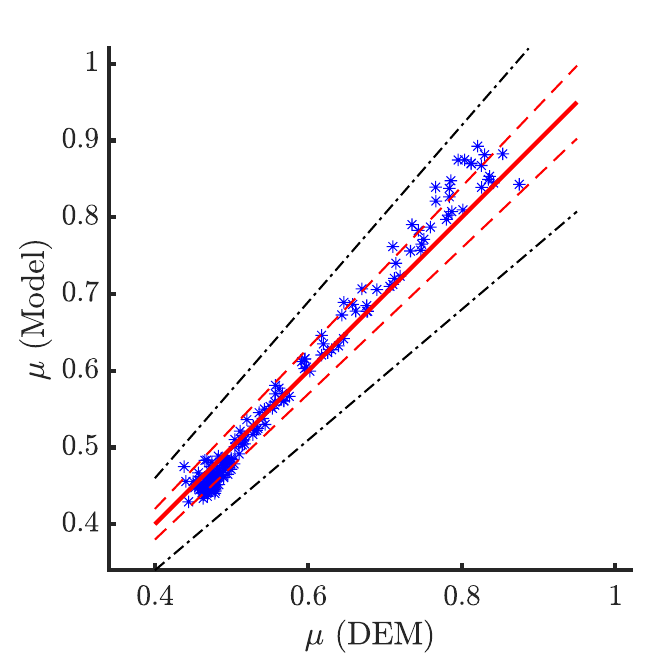}
	\caption{Modified power-law scaling of the stress ratio $\mu$(Model) predictions against numerical experimental data $\mu$(DEM) for all mixtures. The red and black lines indicate the 5\% and 15\% error, respectively.}
	\label{fig:8}
\end{figure}

Our fitting approach enables a well-aligned collapse of data for polydisperse mixtures, as depicted in Figure \ref{fig:7}b. 
The successful collapse of data onto a single line indicates that the stress ratio $\mu$ can be predicted using solely dimensionless parameters - the inertial number $I$, solid concentration $\phi$, and scaled granular temperature $\Theta$.
Furthermore, we deduce that the exponent $p=1/6$ is a universal characteristic, independent of the specific material properties. Despite the inability to fully reconcile the data between monodisperse and polydisperse cases, the necessity of employing the concentration to scale the stress ratio data appears to be unaffected by the components of granular temperature or pressure used (Figures \ref{Supplementary fig:2} and \ref{Supplementary fig:3}). 
The collapse of our monodisperse and polydisperse data has major implications. Before, the stress ratio needed to be defined based on various empirical fits to describe the size and density mixtures. Instead, Figure \ref{fig:7} suggests that a single empirical polynomial law combined with conservation of mass (to describe $\phi$) predicts the stress ratio for all situations investigated. Consequently, the following scaling equation describes the flow rheology:
\begin{eqnarray}
\frac{\mu \Theta ^{1/6}}{\phi ^{2}}=f(I)
\end{eqnarray}
Using this approach, we predict the stress ratio of size, density, friction) mixtures investigated and show that it works very well for $\mu<$0.75 (Figure \ref{fig:8}). The large sensitivity at large inertial numbers with friction that is up to 15\% higher than predicted by Equation \ref{eqn:kamrin_fluidity_fit} could be explored in future work using fitting function of higher order.

In our final analysis, we examine the rheology of both monodisperse and polydisperse grain-size distributions within a variety of flow configurations. These configurations encompass thick flows featuring non-linear shear (evidenced by an exponential velocity profile) and subject to gravity, thin flows devoid of gravity (indicated by a linear velocity profile), and flows moving down a slope (characterized by a concave velocity profile; see Fig.1).

Through the application of the scaling, $\mu\Theta^{p}=f(I)$, the monodisperse and polydisperse data obtained from various flow geometries could not be collapsed upon a single curve (Figure 9a). As a result, the Eq.\ref{eqn:kamrin_fluidity_fit_original}, derived from monodisperse simple shear data, does not provide a satisfactory fit to the data. 

However, We find all data points aligning on a master curve when using the scaling $\mu(I,\Theta,\phi)$, fitted with Eq. \ref{eqn:kamrin_fluidity_fit} (Figure 9b). Therefore, the latter scaling effectively encapsulates the non-local behavior emerging from the spatial diffusion of granular temperature in polydisperse granular mixtures, and allows us to unite with a power-law scaling the rheology of monodisperse and polydisperse grain-size mixtures across flow geometries.

\begin{figure*}[htbp]
	\centering		\includegraphics[width=1\textwidth, angle=0]{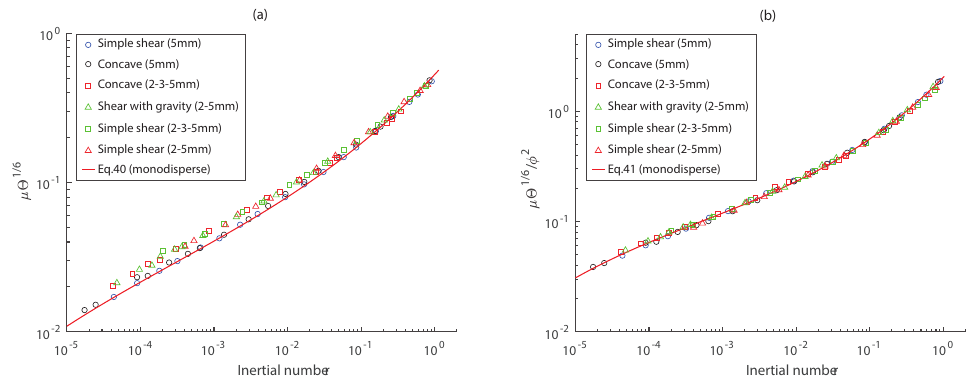}
	\caption{Power-law scaling $\mu \Theta ^{1/6} = f(I)$ (a) and modified power-law scaling $\frac{\mu \Theta ^{1/6}}{\phi^{2}} = f(I)$  (b) for monodisperse and polydisperse mixtures in simple shear, concave and shear with gravity flow geometries. The red line is the fit from the monodisperse simple shear data, given by Eq.\ref{eqn:kamrin_fluidity_fit_original} (a) and Eq.\ref{eqn:kamrin_fluidity_fit} (b).}
	\label{fig:9}
\end{figure*}

\section{\label{sec:Conclusion}Conclusion}
In this work, we have investigated the rheology of shear flows of dense, frictional granular media made of monodisperse, polydisperse and biphasic (friction and density) particles. We observed the occurrence of well-known quasi-static, intermediate and inertial/collisional regimes for all mixtures, with a strong dependence of the critical volume fraction $\phi_{c}$ on the Skewness and Polydispersity and particle-particle friction of the media. We find the inertial number scaling of the stress ratio remains valid as long as the volumetric contribution of the size (using the volume mean diameter $D_{43}$), density and particle friction are accounted for in its definition. In addition, we bind the modified Kinetic Theory \cite{lun1991kinetic,berzi2020extended,garzo1999dense} to the $\mu(I)$-rheology by proposing an updated version of the power-law scaling introduced by Kim and Kamrin \cite{KK2020} to now account for polydisperse size distributions. We have established a new scaling for the stress ratio, which incorporates material properties and hinges on factors such as granular temperature, concentration, and the inertial number. Crucially, it encapsulates both local and non-local rheologies, which manifest as endogeneous velocity fluctuations disperse spatially from areas of high granular temperature towards those with lower temperatures. These findings emphasize the importance of investigating granular temperature and its diffusion, especially in contexts where non-local effects (i.e. large granular temperature gradients) are significant. This would enhance our understanding of polydisperse granular flows across various natural and industrial settings. Moreover, as natural granular media often exhibit rapid spatial changes in response to topographical shifts, further studies focusing on the transient behavior of granular media could provide unifying descriptions of their rheology.\\

\bigskip 
E.C.P.B. was supported by a NERC Independent Research Fellowship (NE/V014242/1). L.F. acknowledges funding from the Royal Society of New Zealand (contracts and MAU1712). 
Financial support was provided by the National Science Foundation grant EAR 1650382 to J.D. 

\bibliographystyle{apsrev4-2}
%\bibliographystyle{unsrt}
%\bibliography{biblio} % replace "references" with the name of your .bib file

%apsrev4-2.bst 2019-01-14 (MD) hand-edited version of apsrev4-1.bst
%Control: key (0)
%Control: author (72) initials jnrlst
%Control: editor formatted (1) identically to author
%Control: production of article title (-1) disabled
%Control: page (0) single
%Control: year (1) truncated
%Control: production of eprint (0) enabled
\begin{thebibliography}{0}%
\makeatletter
\providecommand \@ifxundefined [1]{%
 \@ifx{#1\undefined}
}%
\providecommand \@ifnum [1]{%
 \ifnum #1\expandafter \@firstoftwo
 \else \expandafter \@secondoftwo
 \fi
}%
\providecommand \@ifx [1]{%
 \ifx #1\expandafter \@firstoftwo
 \else \expandafter \@secondoftwo
 \fi
}%
\providecommand \natexlab [1]{#1}%
\providecommand \enquote  [1]{``#1''}%
\providecommand \bibnamefont  [1]{#1}%
\providecommand \bibfnamefont [1]{#1}%
\providecommand \citenamefont [1]{#1}%
\providecommand \href@noop [0]{\@secondoftwo}%
\providecommand \href [0]{\begingroup \@sanitize@url \@href}%
\providecommand \@href[1]{\@@startlink{#1}\@@href}%
\providecommand \@@href[1]{\endgroup#1\@@endlink}%
\providecommand \@sanitize@url [0]{\catcode `\\12\catcode `\$12\catcode
  `\&12\catcode `\#12\catcode `\^12\catcode `\_12\catcode `\%12\relax}%
\providecommand \@@startlink[1]{}%
\providecommand \@@endlink[0]{}%
\providecommand \url  [0]{\begingroup\@sanitize@url \@url }%
\providecommand \@url [1]{\endgroup\@href {#1}{\urlprefix }}%
\providecommand \urlprefix  [0]{URL }%
\providecommand \Eprint [0]{\href }%
\providecommand \doibase [0]{https://doi.org/}%
\providecommand \selectlanguage [0]{\@gobble}%
\providecommand \bibinfo  [0]{\@secondoftwo}%
\providecommand \bibfield  [0]{\@secondoftwo}%
\providecommand \translation [1]{[#1]}%
\providecommand \BibitemOpen [0]{}%
\providecommand \bibitemStop [0]{}%
\providecommand \bibitemNoStop [0]{.\EOS\space}%
\providecommand \EOS [0]{\spacefactor3000\relax}%
\providecommand \BibitemShut  [1]{\csname bibitem#1\endcsname}%
\let\auto@bib@innerbib\@empty
%</preamble>
\end{thebibliography}%


\begin{thebibliography}{34}%
\makeatletter
\providecommand \@ifxundefined [1]{%
 \@ifx{#1\undefined}
}%
\providecommand \@ifnum [1]{%
 \ifnum #1\expandafter \@firstoftwo
 \else \expandafter \@secondoftwo
 \fi
}%
\providecommand \@ifx [1]{%
 \ifx #1\expandafter \@firstoftwo
 \else \expandafter \@secondoftwo
 \fi
}%
\providecommand \natexlab [1]{#1}%
\providecommand \enquote  [1]{``#1''}%
\providecommand \bibnamefont  [1]{#1}%
\providecommand \bibfnamefont [1]{#1}%
\providecommand \citenamefont [1]{#1}%
\providecommand \href@noop [0]{\@secondoftwo}%
\providecommand \href [0]{\begingroup \@sanitize@url \@href}%
\providecommand \@href[1]{\@@startlink{#1}\@@href}%
\providecommand \@@href[1]{\endgroup#1\@@endlink}%
\providecommand \@sanitize@url [0]{\catcode `\\12\catcode `\$12\catcode
  `\&12\catcode `\#12\catcode `\^12\catcode `\_12\catcode `\%12\relax}%
\providecommand \@@startlink[1]{}%
\providecommand \@@endlink[0]{}%
\providecommand \url  [0]{\begingroup\@sanitize@url \@url }%
\providecommand \@url [1]{\endgroup\@href {#1}{\urlprefix }}%
\providecommand \urlprefix  [0]{URL }%
\providecommand \Eprint [0]{\href }%
\providecommand \doibase [0]{https://doi.org/}%
\providecommand \selectlanguage [0]{\@gobble}%
\providecommand \bibinfo  [0]{\@secondoftwo}%
\providecommand \bibfield  [0]{\@secondoftwo}%
\providecommand \translation [1]{[#1]}%
\providecommand \BibitemOpen [0]{}%
\providecommand \bibitemStop [0]{}%
\providecommand \bibitemNoStop [0]{.\EOS\space}%
\providecommand \EOS [0]{\spacefactor3000\relax}%
\providecommand \BibitemShut  [1]{\csname bibitem#1\endcsname}%
\let\auto@bib@innerbib\@empty
%</preamble>
\bibitem [{\citenamefont {Sovilla}\ \emph {et~al.}(2018)\citenamefont
  {Sovilla}, \citenamefont {McElwaine},\ and\ \citenamefont
  {K{\"o}hler}}]{sovilla2018intermittency}%
  \BibitemOpen
  \bibfield  {author} {\bibinfo {author} {\bibfnamefont {B.}~\bibnamefont
  {Sovilla}}, \bibinfo {author} {\bibfnamefont {J.}~\bibnamefont {McElwaine}},\
  and\ \bibinfo {author} {\bibfnamefont {A.}~\bibnamefont {K{\"o}hler}},\
  }\href@noop {} {\bibfield  {journal} {\bibinfo  {journal} {Journal of
  Geophysical Research: Earth Surface}\ }\textbf {\bibinfo {volume} {123}},\
  \bibinfo {pages} {2525} (\bibinfo {year} {2018})}\BibitemShut {NoStop}%
\bibitem [{\citenamefont {Iverson}\ \emph {et~al.}(2011)\citenamefont
  {Iverson}, \citenamefont {Reid}, \citenamefont {Logan}, \citenamefont
  {LaHusen}, \citenamefont {Godt},\ and\ \citenamefont
  {Griswold}}]{iverson2011}%
  \BibitemOpen
  \bibfield  {author} {\bibinfo {author} {\bibfnamefont {R.~M.}\ \bibnamefont
  {Iverson}}, \bibinfo {author} {\bibfnamefont {M.~E.}\ \bibnamefont {Reid}},
  \bibinfo {author} {\bibfnamefont {M.}~\bibnamefont {Logan}}, \bibinfo
  {author} {\bibfnamefont {R.~G.}\ \bibnamefont {LaHusen}}, \bibinfo {author}
  {\bibfnamefont {J.~W.}\ \bibnamefont {Godt}},\ and\ \bibinfo {author}
  {\bibfnamefont {J.~P.}\ \bibnamefont {Griswold}},\ }\href@noop {} {\bibfield
  {journal} {\bibinfo  {journal} {Nature Geoscience}\ }\textbf {\bibinfo
  {volume} {4}},\ \bibinfo {pages} {116} (\bibinfo {year} {2011})}\BibitemShut
  {NoStop}%
\bibitem [{\citenamefont {Hungr}(2006)}]{hungr2006rock}%
  \BibitemOpen
  \bibfield  {author} {\bibinfo {author} {\bibfnamefont {O.}~\bibnamefont
  {Hungr}},\ }in\ \href@noop {} {\emph {\bibinfo {booktitle} {Landslides from
  massive rock slope failure}}}\ (\bibinfo {organization} {Springer},\ \bibinfo
  {year} {2006})\ pp.\ \bibinfo {pages} {243--266}\BibitemShut {NoStop}%
\bibitem [{\citenamefont {Lube}\ \emph {et~al.}(2020)\citenamefont {Lube},
  \citenamefont {Breard}, \citenamefont {Esposti-Ongaro}, \citenamefont
  {Dufek},\ and\ \citenamefont {Brand}}]{lube2020multiphase}%
  \BibitemOpen
  \bibfield  {author} {\bibinfo {author} {\bibfnamefont {G.}~\bibnamefont
  {Lube}}, \bibinfo {author} {\bibfnamefont {E.~C.}\ \bibnamefont {Breard}},
  \bibinfo {author} {\bibfnamefont {T.}~\bibnamefont {Esposti-Ongaro}},
  \bibinfo {author} {\bibfnamefont {J.}~\bibnamefont {Dufek}},\ and\ \bibinfo
  {author} {\bibfnamefont {B.}~\bibnamefont {Brand}},\ }\href@noop {}
  {\bibfield  {journal} {\bibinfo  {journal} {Nature Reviews Earth \&
  Environment}\ }\textbf {\bibinfo {volume} {1}},\ \bibinfo {pages} {348}
  (\bibinfo {year} {2020})}\BibitemShut {NoStop}%
\bibitem [{\citenamefont {Dufek}(2016)}]{dufek2016fluid}%
  \BibitemOpen
  \bibfield  {author} {\bibinfo {author} {\bibfnamefont {J.}~\bibnamefont
  {Dufek}},\ }\href@noop {} {\bibfield  {journal} {\bibinfo  {journal} {Annual
  Review of Fluid Mechanics}\ }\textbf {\bibinfo {volume} {48}},\ \bibinfo
  {pages} {459} (\bibinfo {year} {2016})}\BibitemShut {NoStop}%
\bibitem [{\citenamefont {Cole}\ \emph {et~al.}(2015)\citenamefont {Cole},
  \citenamefont {Neri},\ and\ \citenamefont {Baxter}}]{cole2015hazards}%
  \BibitemOpen
  \bibfield  {author} {\bibinfo {author} {\bibfnamefont {P.~D.}\ \bibnamefont
  {Cole}}, \bibinfo {author} {\bibfnamefont {A.}~\bibnamefont {Neri}},\ and\
  \bibinfo {author} {\bibfnamefont {P.~J.}\ \bibnamefont {Baxter}},\ }in\
  \href@noop {} {\emph {\bibinfo {booktitle} {The encyclopedia of volcanoes}}}\
  (\bibinfo  {publisher} {Elsevier},\ \bibinfo {year} {2015})\ pp.\ \bibinfo
  {pages} {943--956}\BibitemShut {NoStop}%
\bibitem [{\citenamefont {Huber}\ \emph {et~al.}(2020)\citenamefont {Huber},
  \citenamefont {Ojha}, \citenamefont {Lark},\ and\ \citenamefont
  {Head}}]{huber2020physical}%
  \BibitemOpen
  \bibfield  {author} {\bibinfo {author} {\bibfnamefont {C.}~\bibnamefont
  {Huber}}, \bibinfo {author} {\bibfnamefont {L.}~\bibnamefont {Ojha}},
  \bibinfo {author} {\bibfnamefont {L.}~\bibnamefont {Lark}},\ and\ \bibinfo
  {author} {\bibfnamefont {J.~W.}\ \bibnamefont {Head}},\ }\href@noop {}
  {\bibfield  {journal} {\bibinfo  {journal} {Icarus}\ }\textbf {\bibinfo
  {volume} {335}},\ \bibinfo {pages} {113385} (\bibinfo {year}
  {2020})}\BibitemShut {NoStop}%
\bibitem [{\citenamefont {M{\'e}ndez~Harper}\ \emph {et~al.}(2017)\citenamefont
  {M{\'e}ndez~Harper}, \citenamefont {McDonald}, \citenamefont {Dufek},
  \citenamefont {Malaska}, \citenamefont {Burr}, \citenamefont {Hayes},
  \citenamefont {McAdams},\ and\ \citenamefont
  {Wray}}]{mendez2017electrification}%
  \BibitemOpen
  \bibfield  {author} {\bibinfo {author} {\bibfnamefont {J.}~\bibnamefont
  {M{\'e}ndez~Harper}}, \bibinfo {author} {\bibfnamefont {G.}~\bibnamefont
  {McDonald}}, \bibinfo {author} {\bibfnamefont {J.}~\bibnamefont {Dufek}},
  \bibinfo {author} {\bibfnamefont {M.}~\bibnamefont {Malaska}}, \bibinfo
  {author} {\bibfnamefont {D.}~\bibnamefont {Burr}}, \bibinfo {author}
  {\bibfnamefont {A.}~\bibnamefont {Hayes}}, \bibinfo {author} {\bibfnamefont
  {J.}~\bibnamefont {McAdams}},\ and\ \bibinfo {author} {\bibfnamefont
  {J.}~\bibnamefont {Wray}},\ }\href@noop {} {\bibfield  {journal} {\bibinfo
  {journal} {Nature Geoscience}\ }\textbf {\bibinfo {volume} {10}},\ \bibinfo
  {pages} {260} (\bibinfo {year} {2017})}\BibitemShut {NoStop}%
\bibitem [{\citenamefont {Andreotti}\ \emph {et~al.}(2013)\citenamefont
  {Andreotti}, \citenamefont {Forterre},\ and\ \citenamefont
  {Pouliquen}}]{andreotti2013granular}%
  \BibitemOpen
  \bibfield  {author} {\bibinfo {author} {\bibfnamefont {B.}~\bibnamefont
  {Andreotti}}, \bibinfo {author} {\bibfnamefont {Y.}~\bibnamefont
  {Forterre}},\ and\ \bibinfo {author} {\bibfnamefont {O.}~\bibnamefont
  {Pouliquen}},\ }\href@noop {} {\emph {\bibinfo {title} {Granular media:
  between fluid and solid}}}\ (\bibinfo  {publisher} {Cambridge University
  Press},\ \bibinfo {year} {2013})\BibitemShut {NoStop}%
\bibitem [{\citenamefont {Herrmann}\ \emph {et~al.}(2013)\citenamefont
  {Herrmann}, \citenamefont {Hovi},\ and\ \citenamefont
  {Luding}}]{herrmann2013physics}%
  \BibitemOpen
  \bibfield  {author} {\bibinfo {author} {\bibfnamefont {H.~J.}\ \bibnamefont
  {Herrmann}}, \bibinfo {author} {\bibfnamefont {J.-P.}\ \bibnamefont {Hovi}},\
  and\ \bibinfo {author} {\bibfnamefont {S.}~\bibnamefont {Luding}},\
  }\href@noop {} {\emph {\bibinfo {title} {Physics of dry granular media}}},\
  Vol.\ \bibinfo {volume} {350}\ (\bibinfo  {publisher} {Springer Science \&
  Business Media},\ \bibinfo {year} {2013})\BibitemShut {NoStop}%
\bibitem [{\citenamefont {Jop}\ \emph {et~al.}(2006)\citenamefont {Jop},
  \citenamefont {Forterre},\ and\ \citenamefont
  {Pouliquen}}]{jop2006constitutive}%
  \BibitemOpen
  \bibfield  {author} {\bibinfo {author} {\bibfnamefont {P.}~\bibnamefont
  {Jop}}, \bibinfo {author} {\bibfnamefont {Y.}~\bibnamefont {Forterre}},\ and\
  \bibinfo {author} {\bibfnamefont {O.}~\bibnamefont {Pouliquen}},\ }\href@noop
  {} {\bibfield  {journal} {\bibinfo  {journal} {Nature}\ }\textbf {\bibinfo
  {volume} {441}},\ \bibinfo {pages} {727} (\bibinfo {year}
  {2006})}\BibitemShut {NoStop}%
\bibitem [{\citenamefont {gdrmidi@ polytech. univ-mrs. fr http://www. lmgc.
  univ-montp2. fr/MIDI/}(2004)}]{gdr2004dense}%
  \BibitemOpen
  \bibfield  {author} {\bibinfo {author} {\bibfnamefont {G.~M.}\ \bibnamefont
  {gdrmidi@ polytech. univ-mrs. fr http://www. lmgc. univ-montp2. fr/MIDI/}},\
  }\href@noop {} {\bibfield  {journal} {\bibinfo  {journal} {The European
  Physical Journal E}\ }\textbf {\bibinfo {volume} {14}},\ \bibinfo {pages}
  {341} (\bibinfo {year} {2004})}\BibitemShut {NoStop}%
\bibitem [{\citenamefont {Staron}\ \emph {et~al.}(2014)\citenamefont {Staron},
  \citenamefont {Lagr{\'e}e},\ and\ \citenamefont
  {Popinet}}]{staron2014continuum}%
  \BibitemOpen
  \bibfield  {author} {\bibinfo {author} {\bibfnamefont {L.}~\bibnamefont
  {Staron}}, \bibinfo {author} {\bibfnamefont {P.-Y.}\ \bibnamefont
  {Lagr{\'e}e}},\ and\ \bibinfo {author} {\bibfnamefont {S.}~\bibnamefont
  {Popinet}},\ }\href@noop {} {\bibfield  {journal} {\bibinfo  {journal} {The
  European Physical Journal E}\ }\textbf {\bibinfo {volume} {37}},\ \bibinfo
  {pages} {5} (\bibinfo {year} {2014})}\BibitemShut {NoStop}%
\bibitem [{\citenamefont {Fullard}\ \emph {et~al.}(2018)\citenamefont
  {Fullard}, \citenamefont {Davies}, \citenamefont {Neather}, \citenamefont
  {Breard}, \citenamefont {Godfrey},\ and\ \citenamefont
  {Lube}}]{fullard2018testing}%
  \BibitemOpen
  \bibfield  {author} {\bibinfo {author} {\bibfnamefont {L.}~\bibnamefont
  {Fullard}}, \bibinfo {author} {\bibfnamefont {C.}~\bibnamefont {Davies}},
  \bibinfo {author} {\bibfnamefont {A.}~\bibnamefont {Neather}}, \bibinfo
  {author} {\bibfnamefont {E.}~\bibnamefont {Breard}}, \bibinfo {author}
  {\bibfnamefont {A.}~\bibnamefont {Godfrey}},\ and\ \bibinfo {author}
  {\bibfnamefont {G.}~\bibnamefont {Lube}},\ }\href@noop {} {\bibfield
  {journal} {\bibinfo  {journal} {Advanced Powder Technology}\ }\textbf
  {\bibinfo {volume} {29}},\ \bibinfo {pages} {310} (\bibinfo {year}
  {2018})}\BibitemShut {NoStop}%
\bibitem [{\citenamefont {Breard}\ \emph {et~al.}(2019)\citenamefont {Breard},
  \citenamefont {Jones}, \citenamefont {Fullard}, \citenamefont {Lube},
  \citenamefont {Davies},\ and\ \citenamefont {Dufek}}]{breard2019}%
  \BibitemOpen
  \bibfield  {author} {\bibinfo {author} {\bibfnamefont {E.~C.}\ \bibnamefont
  {Breard}}, \bibinfo {author} {\bibfnamefont {J.~R.}\ \bibnamefont {Jones}},
  \bibinfo {author} {\bibfnamefont {L.}~\bibnamefont {Fullard}}, \bibinfo
  {author} {\bibfnamefont {G.}~\bibnamefont {Lube}}, \bibinfo {author}
  {\bibfnamefont {C.}~\bibnamefont {Davies}},\ and\ \bibinfo {author}
  {\bibfnamefont {J.}~\bibnamefont {Dufek}},\ }\href@noop {} {\bibfield
  {journal} {\bibinfo  {journal} {Journal of Geophysical Research: Solid
  Earth}\ }\textbf {\bibinfo {volume} {124}},\ \bibinfo {pages} {1343}
  (\bibinfo {year} {2019})}\BibitemShut {NoStop}%
\bibitem [{\citenamefont {Iverson}(2003)}]{iverson2003debris}%
  \BibitemOpen
  \bibfield  {author} {\bibinfo {author} {\bibfnamefont {R.~M.}\ \bibnamefont
  {Iverson}},\ }\href@noop {} {\bibfield  {journal} {\bibinfo  {journal}
  {Debris-flow hazards mitigation: mechanics, prediction, and assessment}\
  }\textbf {\bibinfo {volume} {1}},\ \bibinfo {pages} {303} (\bibinfo {year}
  {2003})}\BibitemShut {NoStop}%
\bibitem [{\citenamefont {Gu}\ \emph {et~al.}(2017)\citenamefont {Gu},
  \citenamefont {Ozel},\ and\ \citenamefont {Sundaresan}}]{gu2017}%
  \BibitemOpen
  \bibfield  {author} {\bibinfo {author} {\bibfnamefont {Y.}~\bibnamefont
  {Gu}}, \bibinfo {author} {\bibfnamefont {A.}~\bibnamefont {Ozel}},\ and\
  \bibinfo {author} {\bibfnamefont {S.}~\bibnamefont {Sundaresan}},\ }in\
  \href@noop {} {\emph {\bibinfo {booktitle} {2017 AIChE Annual Meeting}}}\
  (\bibinfo {organization} {AIChE},\ \bibinfo {year} {2017})\BibitemShut
  {NoStop}%
\bibitem [{\citenamefont {Kim}\ and\ \citenamefont {Kamrin}(2020)}]{KK2020}%
  \BibitemOpen
  \bibfield  {author} {\bibinfo {author} {\bibfnamefont {S.}~\bibnamefont
  {Kim}}\ and\ \bibinfo {author} {\bibfnamefont {K.}~\bibnamefont {Kamrin}},\
  }\href {https://doi.org/10.1103/PhysRevLett.125.088002} {\bibfield  {journal}
  {\bibinfo  {journal} {Phys. Rev. Lett.}\ }\textbf {\bibinfo {volume} {125}},\
  \bibinfo {pages} {088002} (\bibinfo {year} {2020})}\BibitemShut {NoStop}%
\bibitem [{\citenamefont {Garg}\ \emph {et~al.}(2012)\citenamefont {Garg},
  \citenamefont {Galvin}, \citenamefont {Li},\ and\ \citenamefont
  {Pannala}}]{garg2012}%
  \BibitemOpen
  \bibfield  {author} {\bibinfo {author} {\bibfnamefont {R.}~\bibnamefont
  {Garg}}, \bibinfo {author} {\bibfnamefont {J.}~\bibnamefont {Galvin}},
  \bibinfo {author} {\bibfnamefont {T.}~\bibnamefont {Li}},\ and\ \bibinfo
  {author} {\bibfnamefont {S.}~\bibnamefont {Pannala}},\ }\href@noop {}
  {\bibfield  {journal} {\bibinfo  {journal} {Powder Technology}\ }\textbf
  {\bibinfo {volume} {220}},\ \bibinfo {pages} {122} (\bibinfo {year}
  {2012})}\BibitemShut {NoStop}%
\bibitem [{\citenamefont {Li}\ \emph {et~al.}(2012)\citenamefont {Li},
  \citenamefont {Garg}, \citenamefont {Galvin},\ and\ \citenamefont
  {Pannala}}]{Li2012}%
  \BibitemOpen
  \bibfield  {author} {\bibinfo {author} {\bibfnamefont {T.}~\bibnamefont
  {Li}}, \bibinfo {author} {\bibfnamefont {R.}~\bibnamefont {Garg}}, \bibinfo
  {author} {\bibfnamefont {J.}~\bibnamefont {Galvin}},\ and\ \bibinfo {author}
  {\bibfnamefont {S.}~\bibnamefont {Pannala}},\ }\href@noop {} {\bibfield
  {journal} {\bibinfo  {journal} {Powder Technology}\ }\textbf {\bibinfo
  {volume} {220}},\ \bibinfo {pages} {138} (\bibinfo {year}
  {2012})}\BibitemShut {NoStop}%
\bibitem [{\citenamefont {Zhang}\ and\ \citenamefont
  {Kamrin}(2017)}]{zhang2017microscopic}%
  \BibitemOpen
  \bibfield  {author} {\bibinfo {author} {\bibfnamefont {Q.}~\bibnamefont
  {Zhang}}\ and\ \bibinfo {author} {\bibfnamefont {K.}~\bibnamefont {Kamrin}},\
  }\href@noop {} {\bibfield  {journal} {\bibinfo  {journal} {Physical review
  letters}\ }\textbf {\bibinfo {volume} {118}},\ \bibinfo {pages} {058001}
  (\bibinfo {year} {2017})}\BibitemShut {NoStop}%
\bibitem [{\citenamefont {Weinhart}\ \emph {et~al.}(2016)\citenamefont
  {Weinhart}, \citenamefont {Labra}, \citenamefont {Luding},\ and\
  \citenamefont {Ooi}}]{weinhart2016influence}%
  \BibitemOpen
  \bibfield  {author} {\bibinfo {author} {\bibfnamefont {T.}~\bibnamefont
  {Weinhart}}, \bibinfo {author} {\bibfnamefont {C.}~\bibnamefont {Labra}},
  \bibinfo {author} {\bibfnamefont {S.}~\bibnamefont {Luding}},\ and\ \bibinfo
  {author} {\bibfnamefont {J.~Y.}\ \bibnamefont {Ooi}},\ }\href@noop {}
  {\bibfield  {journal} {\bibinfo  {journal} {Powder technology}\ }\textbf
  {\bibinfo {volume} {293}},\ \bibinfo {pages} {138} (\bibinfo {year}
  {2016})}\BibitemShut {NoStop}%
\bibitem [{\citenamefont {Goldhirsch}(2010)}]{goldhirsch2010stress}%
  \BibitemOpen
  \bibfield  {author} {\bibinfo {author} {\bibfnamefont {I.}~\bibnamefont
  {Goldhirsch}},\ }\href@noop {} {\bibfield  {journal} {\bibinfo  {journal}
  {Granular Matter}\ }\textbf {\bibinfo {volume} {12}},\ \bibinfo {pages} {239}
  (\bibinfo {year} {2010})}\BibitemShut {NoStop}%
\bibitem [{\citenamefont {Weinhart}\ \emph {et~al.}(2013)\citenamefont
  {Weinhart}, \citenamefont {Hartkamp}, \citenamefont {Thornton},\ and\
  \citenamefont {Luding}}]{weinhart2013coarse}%
  \BibitemOpen
  \bibfield  {author} {\bibinfo {author} {\bibfnamefont {T.}~\bibnamefont
  {Weinhart}}, \bibinfo {author} {\bibfnamefont {R.}~\bibnamefont {Hartkamp}},
  \bibinfo {author} {\bibfnamefont {A.~R.}\ \bibnamefont {Thornton}},\ and\
  \bibinfo {author} {\bibfnamefont {S.}~\bibnamefont {Luding}},\ }\href@noop {}
  {\bibfield  {journal} {\bibinfo  {journal} {Physics of fluids}\ }\textbf
  {\bibinfo {volume} {25}},\ \bibinfo {pages} {070605} (\bibinfo {year}
  {2013})}\BibitemShut {NoStop}%
\bibitem [{\citenamefont {Tunuguntla}\ \emph {et~al.}(2017)\citenamefont
  {Tunuguntla}, \citenamefont {Weinhart},\ and\ \citenamefont
  {Thornton}}]{tunuguntla2017comparing}%
  \BibitemOpen
  \bibfield  {author} {\bibinfo {author} {\bibfnamefont {D.~R.}\ \bibnamefont
  {Tunuguntla}}, \bibinfo {author} {\bibfnamefont {T.}~\bibnamefont
  {Weinhart}},\ and\ \bibinfo {author} {\bibfnamefont {A.~R.}\ \bibnamefont
  {Thornton}},\ }\href@noop {} {\bibfield  {journal} {\bibinfo  {journal}
  {Computational Particle Mechanics}\ }\textbf {\bibinfo {volume} {4}},\
  \bibinfo {pages} {387} (\bibinfo {year} {2017})}\BibitemShut {NoStop}%
\bibitem [{\citenamefont {Breard}\ \emph {et~al.}(2022)\citenamefont {Breard},
  \citenamefont {Fullard}, \citenamefont {Dufek}, \citenamefont {Tennenbaum},
  \citenamefont {Fernandez~Nieves},\ and\ \citenamefont
  {Dietiker}}]{breard2022investigating}%
  \BibitemOpen
  \bibfield  {author} {\bibinfo {author} {\bibfnamefont {E.~C.}\ \bibnamefont
  {Breard}}, \bibinfo {author} {\bibfnamefont {L.}~\bibnamefont {Fullard}},
  \bibinfo {author} {\bibfnamefont {J.}~\bibnamefont {Dufek}}, \bibinfo
  {author} {\bibfnamefont {M.}~\bibnamefont {Tennenbaum}}, \bibinfo {author}
  {\bibfnamefont {A.}~\bibnamefont {Fernandez~Nieves}},\ and\ \bibinfo {author}
  {\bibfnamefont {J.~F.}\ \bibnamefont {Dietiker}},\ }\href@noop {} {\bibfield
  {journal} {\bibinfo  {journal} {Granular Matter}\ }\textbf {\bibinfo {volume}
  {24}},\ \bibinfo {pages} {34} (\bibinfo {year} {2022})}\BibitemShut {NoStop}%
\bibitem [{\citenamefont {Fall}\ \emph {et~al.}(2015)\citenamefont {Fall},
  \citenamefont {Ovarlez}, \citenamefont {Hautemayou}, \citenamefont
  {M{\'e}zi{\`e}re}, \citenamefont {Roux},\ and\ \citenamefont
  {Chevoir}}]{fall2015dry}%
  \BibitemOpen
  \bibfield  {author} {\bibinfo {author} {\bibfnamefont {A.}~\bibnamefont
  {Fall}}, \bibinfo {author} {\bibfnamefont {G.}~\bibnamefont {Ovarlez}},
  \bibinfo {author} {\bibfnamefont {D.}~\bibnamefont {Hautemayou}}, \bibinfo
  {author} {\bibfnamefont {C.}~\bibnamefont {M{\'e}zi{\`e}re}}, \bibinfo
  {author} {\bibfnamefont {J.-N.}\ \bibnamefont {Roux}},\ and\ \bibinfo
  {author} {\bibfnamefont {F.}~\bibnamefont {Chevoir}},\ }\href@noop {}
  {\bibfield  {journal} {\bibinfo  {journal} {Journal of rheology}\ }\textbf
  {\bibinfo {volume} {59}},\ \bibinfo {pages} {1065} (\bibinfo {year}
  {2015})}\BibitemShut {NoStop}%
\bibitem [{\citenamefont {Berzi}\ \emph {et~al.}(2020)\citenamefont {Berzi},
  \citenamefont {Jenkins},\ and\ \citenamefont {Richard}}]{berzi2020extended}%
  \BibitemOpen
  \bibfield  {author} {\bibinfo {author} {\bibfnamefont {D.}~\bibnamefont
  {Berzi}}, \bibinfo {author} {\bibfnamefont {J.~T.}\ \bibnamefont {Jenkins}},\
  and\ \bibinfo {author} {\bibfnamefont {P.}~\bibnamefont {Richard}},\
  }\href@noop {} {\bibfield  {journal} {\bibinfo  {journal} {Journal of Fluid
  Mechanics}\ }\textbf {\bibinfo {volume} {885}},\ \bibinfo {pages} {A27}
  (\bibinfo {year} {2020})}\BibitemShut {NoStop}%
\bibitem [{\citenamefont {Lun}\ \emph {et~al.}(1984)\citenamefont {Lun},
  \citenamefont {Savage}, \citenamefont {Jeffrey},\ and\ \citenamefont
  {Chepurniy}}]{lun1984kinetic}%
  \BibitemOpen
  \bibfield  {author} {\bibinfo {author} {\bibfnamefont {C.~K.}\ \bibnamefont
  {Lun}}, \bibinfo {author} {\bibfnamefont {S.~B.}\ \bibnamefont {Savage}},
  \bibinfo {author} {\bibfnamefont {D.}~\bibnamefont {Jeffrey}},\ and\ \bibinfo
  {author} {\bibfnamefont {N.}~\bibnamefont {Chepurniy}},\ }\href@noop {}
  {\bibfield  {journal} {\bibinfo  {journal} {Journal of fluid mechanics}\
  }\textbf {\bibinfo {volume} {140}},\ \bibinfo {pages} {223} (\bibinfo {year}
  {1984})}\BibitemShut {NoStop}%
\bibitem [{\citenamefont {Garz{\'o}}\ and\ \citenamefont
  {Dufty}(1999)}]{garzo1999dense}%
  \BibitemOpen
  \bibfield  {author} {\bibinfo {author} {\bibfnamefont {V.}~\bibnamefont
  {Garz{\'o}}}\ and\ \bibinfo {author} {\bibfnamefont {J.}~\bibnamefont
  {Dufty}},\ }\href@noop {} {\bibfield  {journal} {\bibinfo  {journal}
  {Physical Review E}\ }\textbf {\bibinfo {volume} {59}},\ \bibinfo {pages}
  {5895} (\bibinfo {year} {1999})}\BibitemShut {NoStop}%
\bibitem [{\citenamefont {Jenkins}\ and\ \citenamefont
  {Savage}(1983)}]{jenkins1983theory}%
  \BibitemOpen
  \bibfield  {author} {\bibinfo {author} {\bibfnamefont {J.~T.}\ \bibnamefont
  {Jenkins}}\ and\ \bibinfo {author} {\bibfnamefont {S.~B.}\ \bibnamefont
  {Savage}},\ }\href@noop {} {\bibfield  {journal} {\bibinfo  {journal}
  {Journal of fluid mechanics}\ }\textbf {\bibinfo {volume} {130}},\ \bibinfo
  {pages} {187} (\bibinfo {year} {1983})}\BibitemShut {NoStop}%
\bibitem [{\citenamefont {Kim}\ and\ \citenamefont {Kamrin}(2023)}]{kim2023}%
  \BibitemOpen
  \bibfield  {author} {\bibinfo {author} {\bibfnamefont {S.}~\bibnamefont
  {Kim}}\ and\ \bibinfo {author} {\bibfnamefont {K.}~\bibnamefont {Kamrin}},\
  }\href@noop {} {\bibfield  {journal} {\bibinfo  {journal} {Frontiers in
  Physics}\ }\textbf {\bibinfo {volume} {11}},\ \bibinfo {pages} {1092233}
  (\bibinfo {year} {2023})}\BibitemShut {NoStop}%
\bibitem [{\citenamefont {Zhu}\ \emph {et~al.}(2023)\citenamefont {Zhu},
  \citenamefont {Huang},\ and\ \citenamefont {Sun}}]{zhu2023granular}%
  \BibitemOpen
  \bibfield  {author} {\bibinfo {author} {\bibfnamefont {C.}~\bibnamefont
  {Zhu}}, \bibinfo {author} {\bibfnamefont {Y.}~\bibnamefont {Huang}},\ and\
  \bibinfo {author} {\bibfnamefont {J.}~\bibnamefont {Sun}},\ }\href@noop {}
  {\bibfield  {journal} {\bibinfo  {journal} {Computers and Geotechnics}\
  }\textbf {\bibinfo {volume} {154}},\ \bibinfo {pages} {105115} (\bibinfo
  {year} {2023})}\BibitemShut {NoStop}%
\bibitem [{\citenamefont {Lun}(1991)}]{lun1991kinetic}%
  \BibitemOpen
  \bibfield  {author} {\bibinfo {author} {\bibfnamefont {C.~K.}\ \bibnamefont
  {Lun}},\ }\href@noop {} {\bibfield  {journal} {\bibinfo  {journal} {Journal
  of fluid mechanics}\ }\textbf {\bibinfo {volume} {233}},\ \bibinfo {pages}
  {539} (\bibinfo {year} {1991})}\BibitemShut {NoStop}%
\end{thebibliography}

%

%merlin.mbs apsrev4-1.bst 2010-07-25 4.21a (PWD, AO, DPC) hacked
%Control: key (0)
%Control: author (0) dotless jnrlst
%Control: editor formatted (1) identically to author
%Control: production of article title (0) allowed
%Control: page (1) range
%Control: year (0) verbatim
%Control: production of eprint (0) enabled

\newpage
\section{Supplementary Figures}
\setcounter{figure}{0}  
\renewcommand{\thefigure}{S\arabic{figure}}  

\begin{figure*}[htbp]
	\centering
	\includegraphics[width=0.6\textwidth, angle=0]{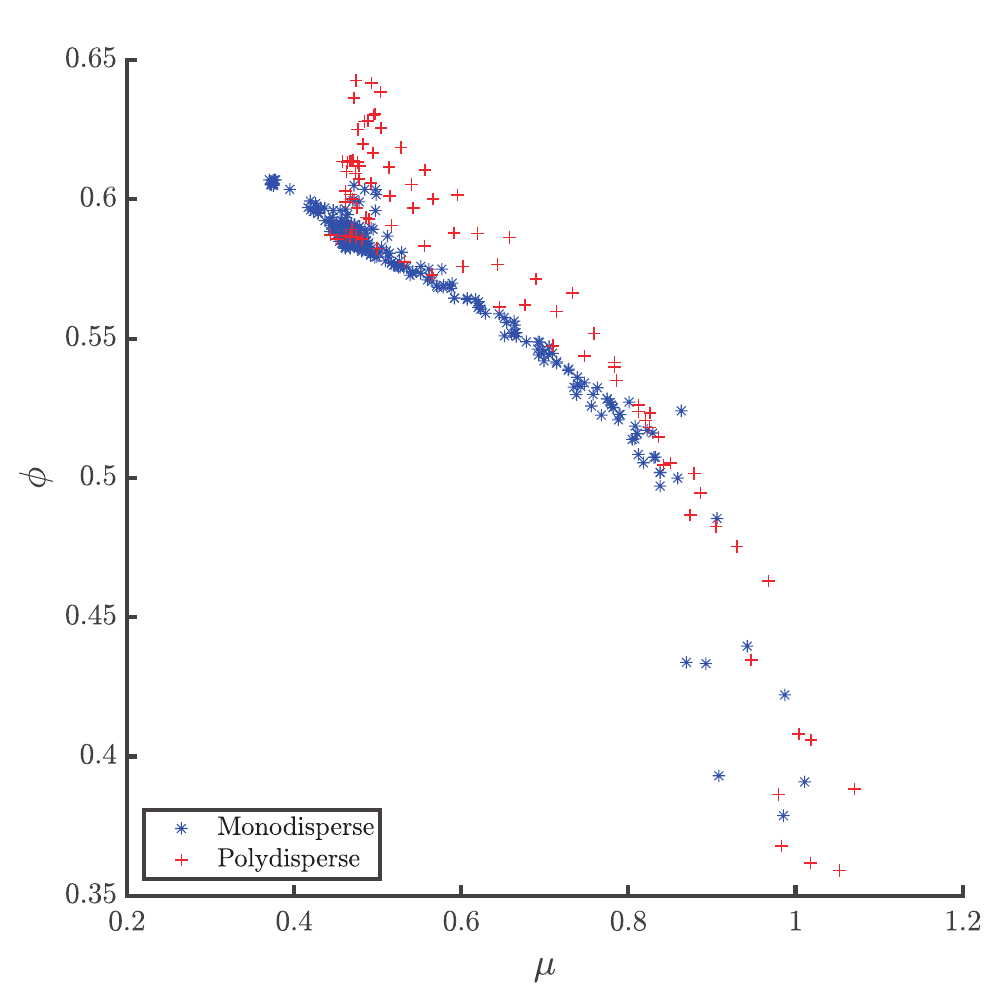}
	\caption{Power-law scaling $\frac{\mu \Theta ^{1/6}}{\phi^{2}} = f(I)$ for monodisperse and density mixtures (a) and size mixtures (b) using the x-component of granular temperature and y-component of the pressure similar to \cite{KK2020}}.
	\label{Supplementary fig:1}
\end{figure*}

\begin{figure*}[htbp]
	\centering		\includegraphics[width=0.95\textwidth, angle=0]{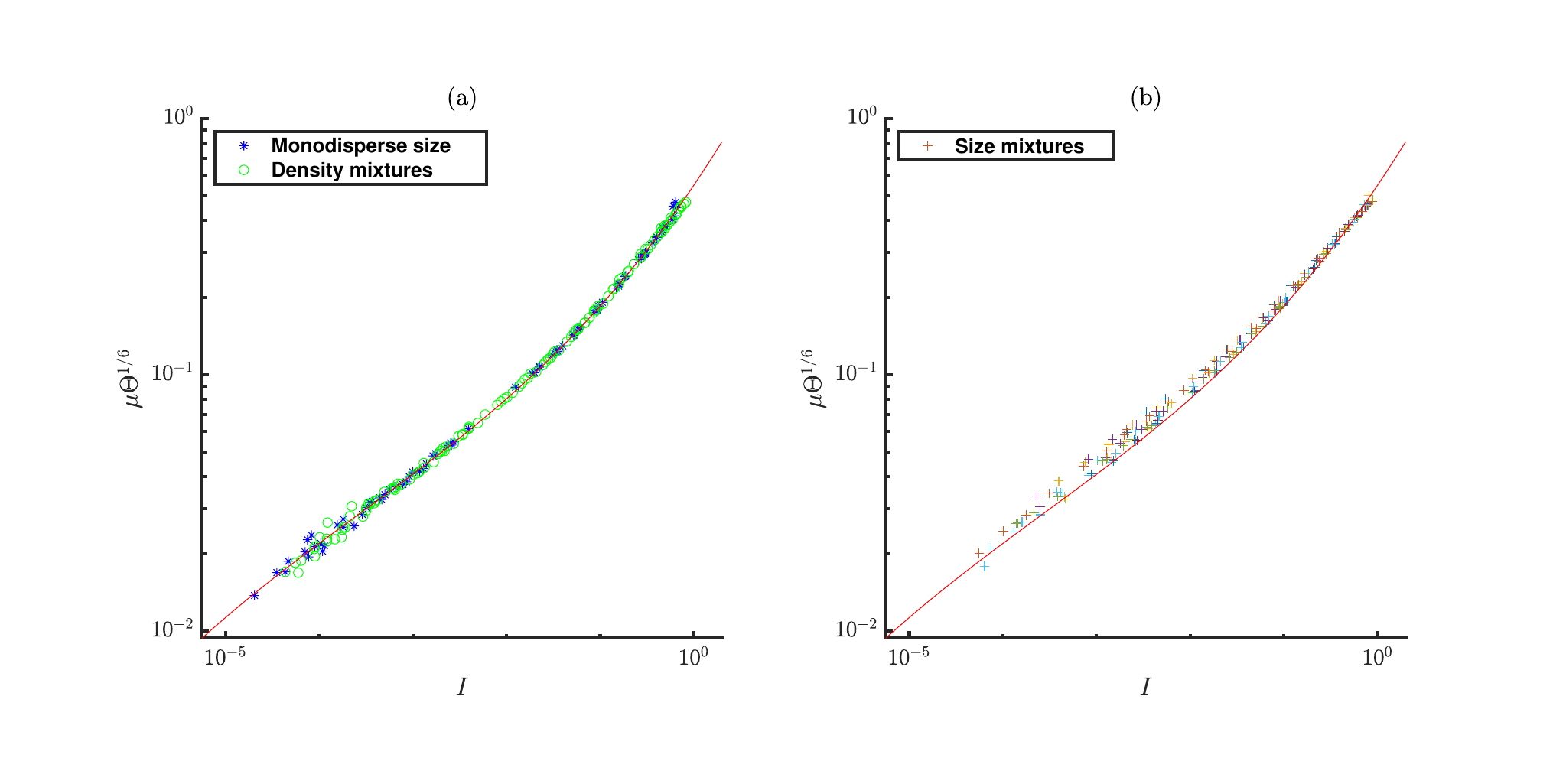}
	\caption{Modified Power-law scaling $\frac{\mu \Theta ^{1/6}}{\phi^{2}} = f(I)$ for monodisperse and density mixtures (a) and size mixtures (b) using the x-component of granular temperature and y-component of the pressure similar to \cite{KK2020}. The red line is the same on both plots.}
	\label{Supplementary fig:2}
\end{figure*}

\begin{figure*}[htbp]
	\centering		\includegraphics[width=0.95\textwidth, angle=0]{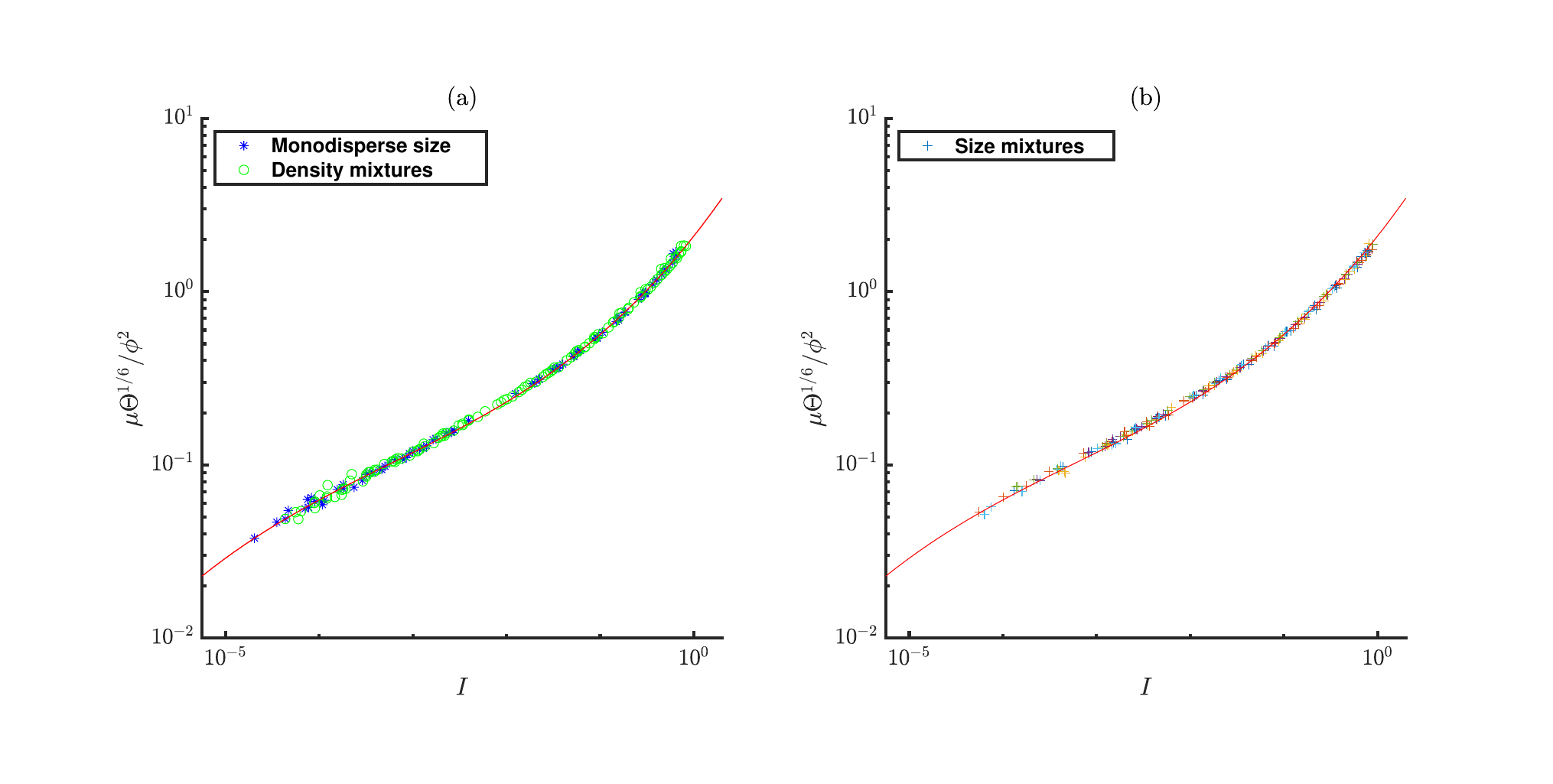}
	\caption{Modified Power-law scaling $\frac{\mu \Theta ^{1/6}}{\phi^{2}} = f(I)$ for monodisperse and density mixtures (a) and size mixtures (b) using the x-component of granular temperature and y-component of the pressure similar to \cite{KK2020}. The red line is the same on both plots.}.
	\label{Supplementary fig:3}
\end{figure*}

\end{document}